\documentclass[fleqn,usenatbib]{mnras}

\usepackage{newtxtext,newtxmath}

\usepackage[T1]{fontenc}

\DeclareRobustCommand{\VAN}[3]{#2}
\let\VANthebibliography\thebibliography
\def\thebibliography{\DeclareRobustCommand{\VAN}[3]{##3}\VANthebibliography}
\defcitealias{Whitworth2023}{WH23}

\usepackage{graphicx}	
\usepackage{amsmath}	
\usepackage{caption}
\usepackage{subcaption}

\title[Magnetic fields and cold gas formation]{A bottleneck for star formation: the importance of magnetic fields during the formation of cold gas in galaxies.}

\author[McGuiness, Smith, and Whitworth]{
Ryan McGuiness,$^{1}$\thanks{E-mail: ryanmcguiness54@outlook.com}
Rowan J. Smith,$^{1}\thanks{E-mail: rjs22@st-andrews.ac.uk}$
David Whitworth$^{2,3}$
\\
$^{1}$SUPA, School of Physics and Astronomy, University of St Andrews, North Haugh, St Andrews, KY16 9SS, UK\\
$^{2}$Universidad Nacional Aut\'onoma de M\'exico, Instituto de Radioastronom\'ia y Astrof\'isica, Antigua Carretera a P\'atzcuaro 8701, Ex-Hda. \\
San Jos\'e de la Huerta, 58089 Morelia, Michoac\'an, M\'exico\\
$^{3}$ENS de Lyon, CRAL UMR5574, Universite Claude Bernard Lyon 1, CNRS, Lyon 69007, France\\
}
\date{Accepted XXX. Received YYY; in original form ZZZ}

\pubyear{\the\year{}}

\begin{document}
\label{firstpage}
\pagerange{\pageref{firstpage}--\pageref{lastpage}}

\maketitle

\begin{abstract}
Using a high-resolution simulation of a dwarf galaxy, we quantify the energetic importance of magnetic fields within the different phases of its interstellar medium (ISM) on parsec scales. We show that, whilst overall the magnetic field is only energetically dominant for a small fraction of the ISM, 
it becomes important in the thermally unstable regime ($45.2\%$ of the mass is magnetically dominant), and dominates in the cold neutral medium ($66.1\%$ of the mass). In the molecular gas, the magnetic field dominates more of the total mass budget ($39.8\%$) than thermal energy, turbulent kinetic energy, or gas self-gravitating potential energy. However, much of this mass will be CO-dark. This suggests that magnetic forces are non-negligible during the formation of cold dense gas, which will slow its collapse and lead to an increase in the fraction of cold atomic and molecular gas in the ISM. Consequently, star-forming clouds may be surrounded by a larger reservoir of cold gas than would otherwise be expected. 
\end{abstract}

\begin{keywords}
galaxies: ISM -- ISM: clouds -- ISM: structure -- stars: formation
\end{keywords}



\section{Introduction}
\label{intro}

The formation of stars, and the evolution of galaxies are both controlled by the continuous cycle of gas between different states within the interstellar medium (ISM). Gas transitions from hot to warm to cold gas where stars form. The feedback from these stars then injects energy and momentum into their natal gas clouds, destroying them and restarting the process \citep[e.g.][]{Draine2011_book,Angles2017,Peroux2020}. The formation of stars within molecular clouds \citep[e.g.][]{Lada2003,Bergin2007,Chevance2023,Schinnerer2024} and the importance of stellar feedback \citep[e.g.][]{Hopkins2012,Dale2015,Padoan2016,Grisdale2017} is much studied. However, our understanding of the formation of cold molecular clouds themselves is less developed, particularly in the context of magnetic fields. This cloud formation process will act as a limiting step in the formation of stars, as without cold dense gas there can be no star-forming cores.

The interstellar medium is comprised of various phases, each with a different temperature and density. These phases—hot, warm, and cold gas—exist in a state of pressure equilibrium \citep{Field1969,Draine1978,Wolfire1995,Wolfire2003}. For cloud formation, the most relevant phase transition is that from the Warm Neutral Medium (WNM) to the Cold Neutral Medium (CNM). As gas within the WNM cools and increases in density, it reaches a thermal instability in which small perturbations cause a rapid decrease in temperature and increase in density. If this occurs, the gas then transitions to become part of the CNM. Within this cold dense gas, molecular hydrogen begins to form as the gas becomes dense enough to self-shield from the interstellar radiation field. Recent numerical simulations have shown emerging evidence that the resulting molecular gas fraction is sensitive to the presence of magnetic fields \citep{Girichidis2018,Ntormousi2017,Kim2021,Whitworth2023}, though the reasons for this are still undetermined.

Models and observations of the magnetic energy densities within spiral galaxies~\citep{Beck1996} find them to be globally in equipartition with the turbulent energy densities within the gas. Recently \citet{Seta2025} used Zeeman magnetic field measurements from the literature and supplemented these with the magnetic field estimates in the ionised ISM using pulsar observations, to show that along sight-lines in the Milky Way turbulent and magnetic energies were also directly proportional to each other. Furthermore, due to magnetic flux conservation, magnetic field strength increases with gas density~\cite[]{Crutcher2010,Whitworth2025b} (unless mass flow is purely along field lines), and so field magnitudes are higher in molecular clouds than in the diffuse ISM. Thus, we expect that magnetic fields must play a role in the energy balance of the ISM and the formation of clouds \citep{Hennebelle2019,Pattle2023}. 

Another indicator of the importance of magnetic fields within the ISM is the field orientation with respect to gas structures. Within diffuse gas, the magnetic field is preferentially parallel to the cold HI atomic gas \cite[]{Clark2014,Clark2019,KimD2023}. In regions of increased density, such as in molecular clouds, this orientation changes to being preferentially perpendicular~\citep[]{Soler2013,PlanckCollab2016,Heyer2020}. This has been interpreted as the magnetic field becoming sub-dominant in denser regions allowing the field to flip as the gas undergoes gravitational collapse and/or convergent flows \citep{Soler2017,Seifried2020}.

Within molecular clouds, magnetic fields are commonly believed to be less important as the field becomes supercritical compared to the global gravitational energy of the molecular cloud \citep{Crutcher2012}. However, magnetic fields may still play a role in the creation of dense gas filaments \citep{Inoue2013,Hennebelle2013a,Xu2019,Kong2024,Abe2021} and there is conflicting evidence of how they may affect the mass spectrum of gas cores formed by fragmentation within the clouds \citep{Palau2015,Palau2021,Weis2024,Klos2025}.

The uncertainties in observations of magnetic fields are large, and any measurements may be affected by, amongst others, line-of-sight effects \citep{Pattle2023}. Therefore, numerical simulations are a necessary supplement to gain a fuller picture of their importance in the ISM. One approach to quantify this is to consider the energy balance within the gas. \cite{Ibanez2022} used an energy analysis of magnetised star forming clouds to show that cloud centres are gravitationally dominated, with gravitational energy exceeding internal, kinetic, or magnetic energy, but cloud envelopes are magnetically supported. Similarly, \citep{Ganguly2024} used an energy balance analysis to investigate the importance of tidal forces in molecular clouds. However, these previous analysis have focused on molecular clouds rather than all the phases of the ISM. Furthermore, \citet{Whitworth2025b} have shown that in order to accurately capture the relation between magnetic field strength and density in simulations the field needs to have evolved naturally with the gas over multiple rotations so it can trace both the dense and diffuse gas in a rotationally supported galactic environment 
- something that is not possible in the previous analyses, which used stratified box simulations. Simulating dwarf galaxies allows for the modelling of a complete galaxy in a steady star formation state, with a full ISM cycle between gas phases, and long (of the order $\sim 1$\,Gyr) evolutionary times allowing for multiple rotations. This is computationally difficult for larger systems. 

Using such a simulation, we seek in this paper to investigate the internal energy balance of the ISM on scales of order a parsec and consequently its importance in regulating the dynamical balance of gas in the interstellar medium of galaxies.

\section{Methods}

\subsection{Numerical Method}
\label{numerics}

A full description of our setup can be found in \citep{Whitworth2023}, hereafter referred to as \citetalias{Whitworth2023}, however, for completeness, we summarise the numerical techniques here. We use the moving mesh code \textsc{Arepo} \citep{Springel2010,Pakmor2011}, which solves the ideal magnetohydrodynamic (MHD) equations using an unstructured Voronoi mesh. To manage divergence errors, we employ the \citet{Powell1999} divergence cleaning technique, paired with an HLLD (Harten-Lax-van Leer Discontinuities) Riemann solver \citep{Miyoshi2005}.

To accurately capture the phases of the ISM we employ a time-dependent, non-equilibrium chemical network based on the work of \cite{Gong2017} that also includes the effects of cosmic ray ionisation. The network tracks nine non-equilibrium species live within the simulation with an additional eight species derived from conservation laws. We assume an ambient interstellar radiation field and use the \textsc{TreeCol} algorithm \citep{Clark2012} to calculate the gas shielding to a radius of 30\,pc. 

We model star formation using an accreting sink particle \citep{Bate1995} approach (see \citealt{Tress2020} for a detailed discussion on our implementation). To form a sink, we check to see if the gas is infalling with negative divergence in both the velocity and acceleration, only allowing sinks to form in \textit{gravitationally bound} gas with density above $n=1000$ cm$^{-3}$. Our sinks have an accretion radius of $1.75$\,pc, within which other sinks cannot form during one local free fall time, meaning that they correspond to star-forming clumps. 

A star formation efficiency within the sink particle of 10\% is assumed, meaning that 10\% of a sinks mass is considered stellar, and the rest is considered to be unresolved gas. The sinks are, therefore, associated with discrete stellar populations as described in \citet{Sormani2017} and generate individual supernovae explosions (SNe) that return energy and momentum back to the ISM when stars of mass 8\,M$_\odot$ or greater reach the end of their lifetime \citep{Tress2020}. We do not include photoionising feedback from individual HII regions in this setup and consequently refrain from discussing the Hot and Warm Ionised Medium (HIM, WIM) in this work. Sinks that will never go supernovae, or have completed their last SNe, return their remaining gas mass to the surrounding ISM. They then get transformed into collisionless $N$-body star particles.

Gravity is captured using the \textsc{Arepo} tree. We use an adaptive softening for the gas cells that matches their size, and fixed softening lengths of 64\,pc for the dark matter, and 2\,pc for the sinks particles and collisionless star particles. 

\subsection{Dwarf Galaxy Simulations}
\label{models}

We perform our analysis upon a dwarf galaxy model simulated using the above techniques. As mentioned in Section \ref{intro}, dwarf galaxies are advantageous for computational reasons, since an entire galaxy can be simulated at high enough resolution to fully resolve the cold phase of the ISM everywhere in the galaxy. This reduction in computational cost allows for the inclusion of detailed chemical gas physics, magnetic fields and extended evolutionary timescales. 

Our fiducial model, MHD\_SOL, is a completely new model run for this analysis that is based on the dwarf galaxy simulations presented in \citetalias{Whitworth2023}. The new model is identical to MHD\_SAT from that paper, except that we set the metallicity as $Z = Z_\odot$, the dust-to-gas ratio (DG$_r$) to 1:100, the cosmic-ray ionisation rate $\zeta_{\rm H}$ to 3 $\times 10^{-17}$, and set a constant far-UV (FUV) background field $G$ to 1.7. These values are in line with the values accepted for the solar neighbourhood and so should be more representative of the phase balance of the ISM in nearby clouds of the Milky Way. Table \ref{parameters} summarises the initial values used for our fiducial model, plus a comparison model that we later use in Section \ref{sec:metallicity}.

\begin{figure}
    \centering
    \includegraphics[width=\linewidth]{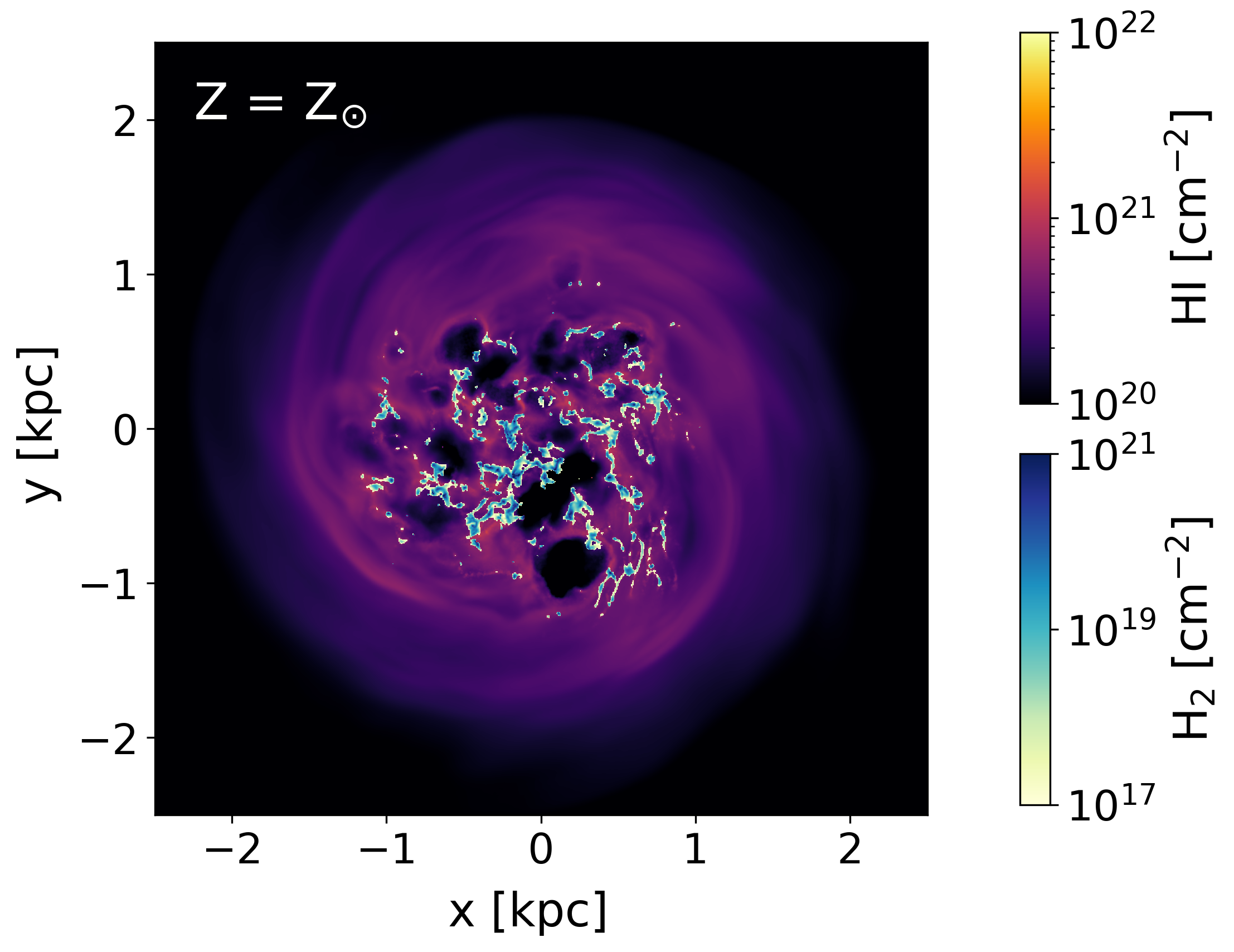}
    \caption{Projection of the HI and H$_2$ surface density for our fiducial model. The molecular gas chiefly lies in the inner galaxy.} 
    \label{fig:SD}
\end{figure}

\begin{figure}
    \centering
    \includegraphics[width=\linewidth]{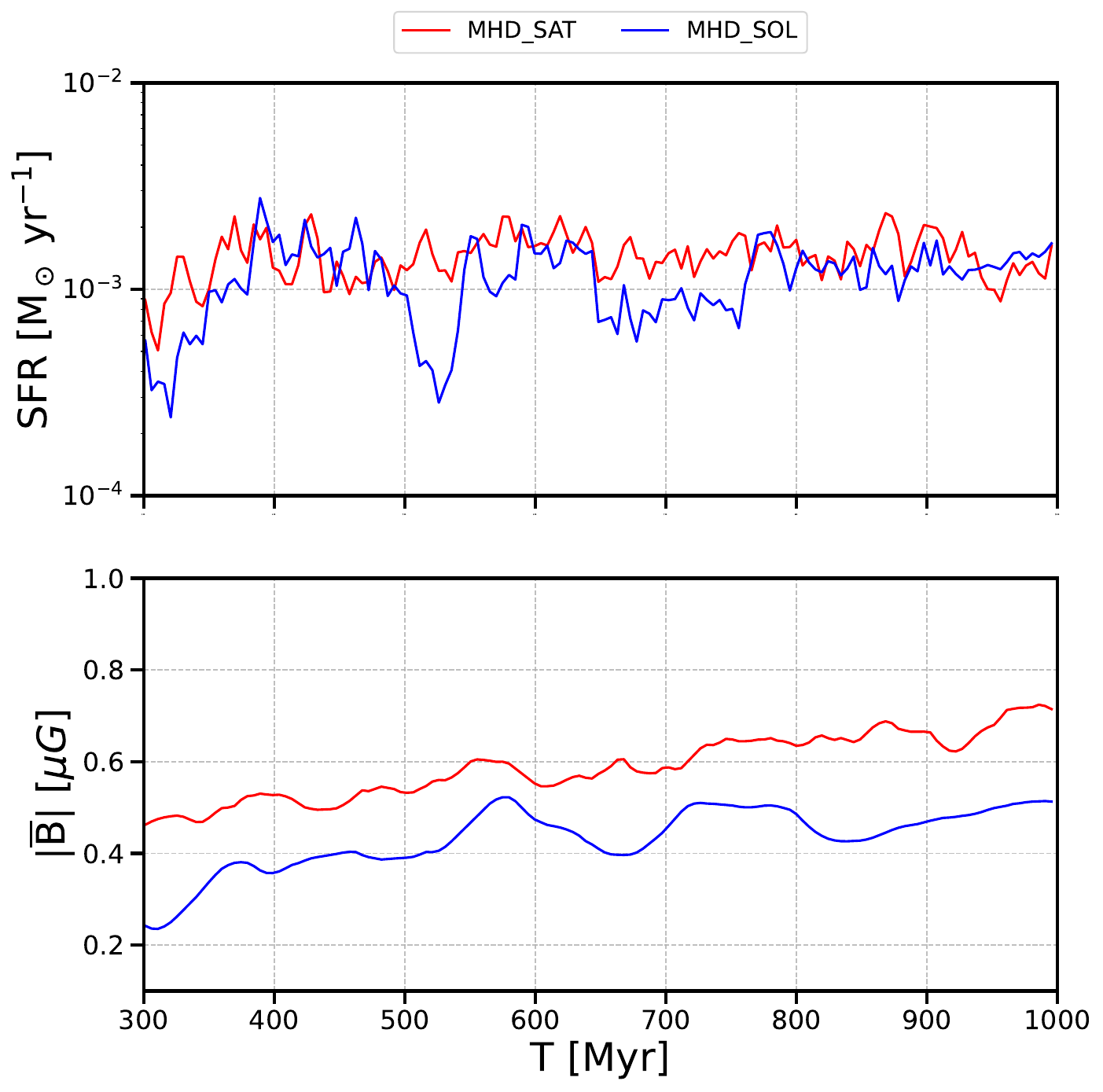}
    \caption{The star formation rate and volume weighted magnetic field strength of our fiducial model MHD\_SOL (blue)  over the steady state period, as described in \protect \citetalias{Whitworth2023}.  The model has reached a steady star formation rate at the time of analysis (1 Gyr). A second low metallicity model (MHD\_SAT in red) is shown for later comparison in Section \ref{sec:metallicity}.
    }
    \label{fig:SFR}
\end{figure}

We use a live dark matter halo initialised with a spheroidal \citealt{Hernquist1990} profile, and a gas disc of initial mass $M_{\rm{disc}} = 8\times10^7$ M$_{\odot}$, vertical scale height $h_z = 0.35$ kpc, and scale length $h_R = 0.82$ kpc. The model does not initially contain a stellar population, but we allow this to form naturally from the gas disc as the galaxy evolves. The simulations are analysed at a time of $1$\,Gyr, when they are fully relaxed, and both star formation and chemical evolution are in a steady state. Figure \ref{fig:SD} shows a face-on view of the dwarf galaxy model at the time of our analysis. (We note that the volume-weighted magnetic field strength, Figure \ref{fig:SFR}, shows the field is likely still growing, possibly driven by the slow build-up of the large-scale dynamo \citep{Korpi2024}. As this is only a minimal growth, $\sim 0.1 \mu$G over 500\,Myr, for our purposes the field can be considered steady state.)

The model is initiated with a poloidal seed field at $10^{-2} \mu$G. As in \citetalias{Whitworth2023}, the initial field strength and orientation have no effect on the final star formation rates or the chemical evolutionary state, and the model has reached a consistent magnetic field strength and star formation rate as shown in Figure \ref{fig:SFR}. 

\begin{figure}
    \centering
    \includegraphics[width=1.0\linewidth]{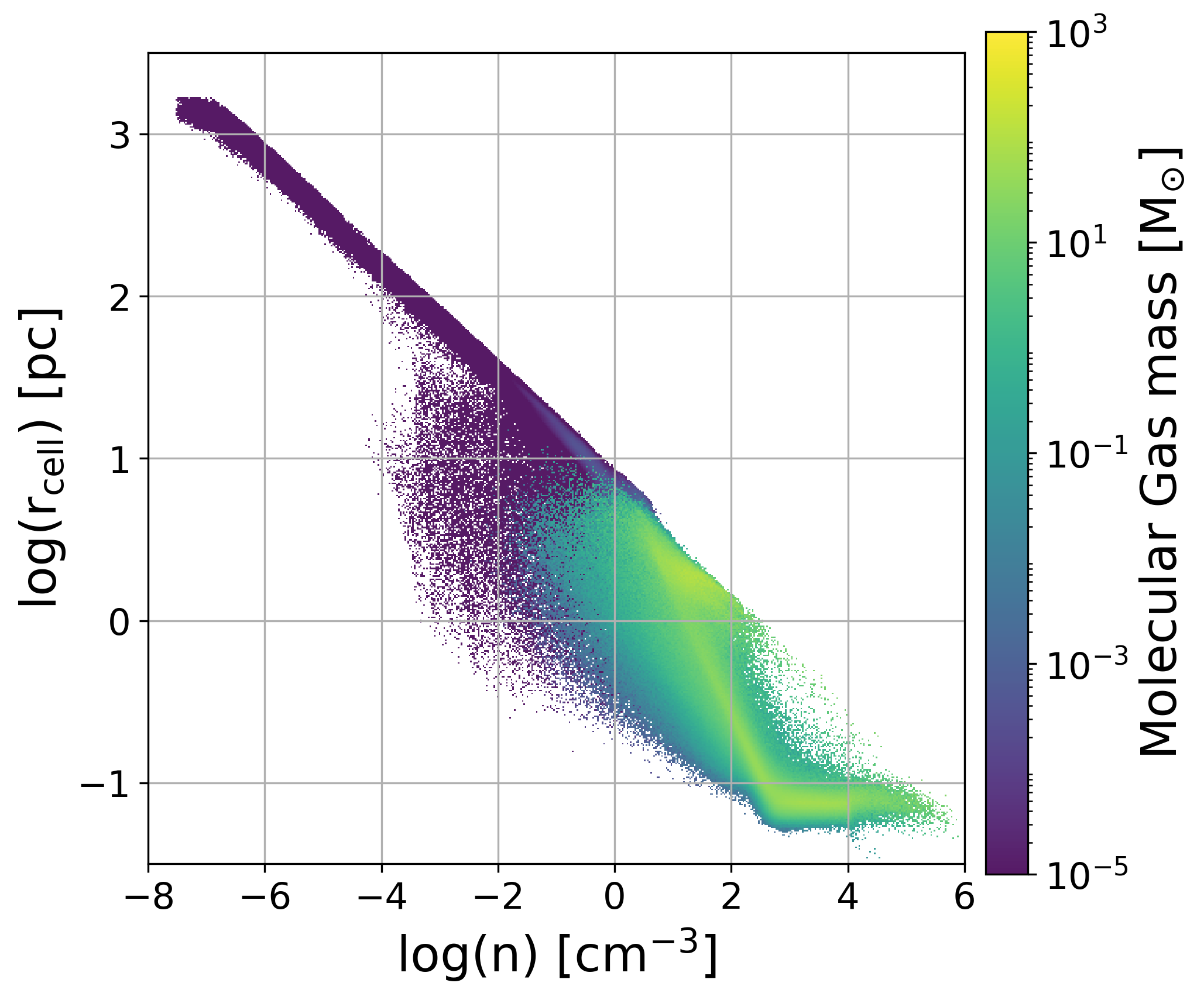}
    \caption{Cell radius as a function of number density for our fiducial model, where the radius is calculated as that of a sphere of the same volume as the cell. All logs are in base 10. The resolution is sub-parsec for all densities greater than 100 cm$^{-3}$.}
    \label{fig:resolution_plot}
\end{figure}

To ensure that the gas is resolved down to the cold clump scale, the target gas mass of each cell is set to 50\,M$_\odot$, meaning the refinement/derefinement routine acts to keep the cell mass within a factor of 2 of this value. On top of this, we set an additional `Jeans refinement' requirement that ensures there is no artificial fragmentation by requiring the local Jeans Length to be resolved by at least 8 resolution elements \citep{Jeans1902,Truelove1997,Greif11}. This stringent second condition means that the cell radii $r_{\rm{cell}}$ reach sub-parsec scales at number densities of $n \sim 100 \rm cm^{-3}$. Figure \ref{fig:resolution_plot} shows our cell resolution, $r_{cell}$, as a function of density. Throughout the paper, we adopt the convention that `log' is equivalent to log$_{10}$. Above the number densities of $n \sim 10 \rm cm^{-3}$ we see a sharp drop in $r_{\rm{cell}}$ where the Jeans refinement criterion is activated due to the high densities and low temperatures.

\begin{table}
\centering
\begin{tabular}{c c c c c c c c c}
 \hline
  Model & Z & DG$_r$ &  $\zeta_{\rm H}$ & $G$ & $\log_{10} $B$_0$\\
            &  ($Z_\odot$) &   & (s$^{-1}$) &  ($G_0$) &  ($\mu$G)  \\
 \hline
 MHD\_SOL & 1.00 & 1.0 & 3 $\times 10^{-17}$ & 1.70 & -2 \\
 MHD\_SAT & 0.10 & 0.1 & 3 $\times 10^{-18}$ & 0.17 & -2 \\

\hline
\end{tabular}
\caption{Initial conditions for our fiducial dwarf galaxy model MHD\_SOL and a comparison model discussed later in the paper. We show metallicity $Z$, dust-to-gas ratio DG$_r$, relative to the value in solar metallicity gas, cosmic ionisation rate $\zeta_{\rm H}$, UV field strength $G$, and magnetic seed field strength $B_0$.}
\label{parameters}
\end{table}

\begin{figure}
    \centering
    \includegraphics[width=1.0\linewidth]{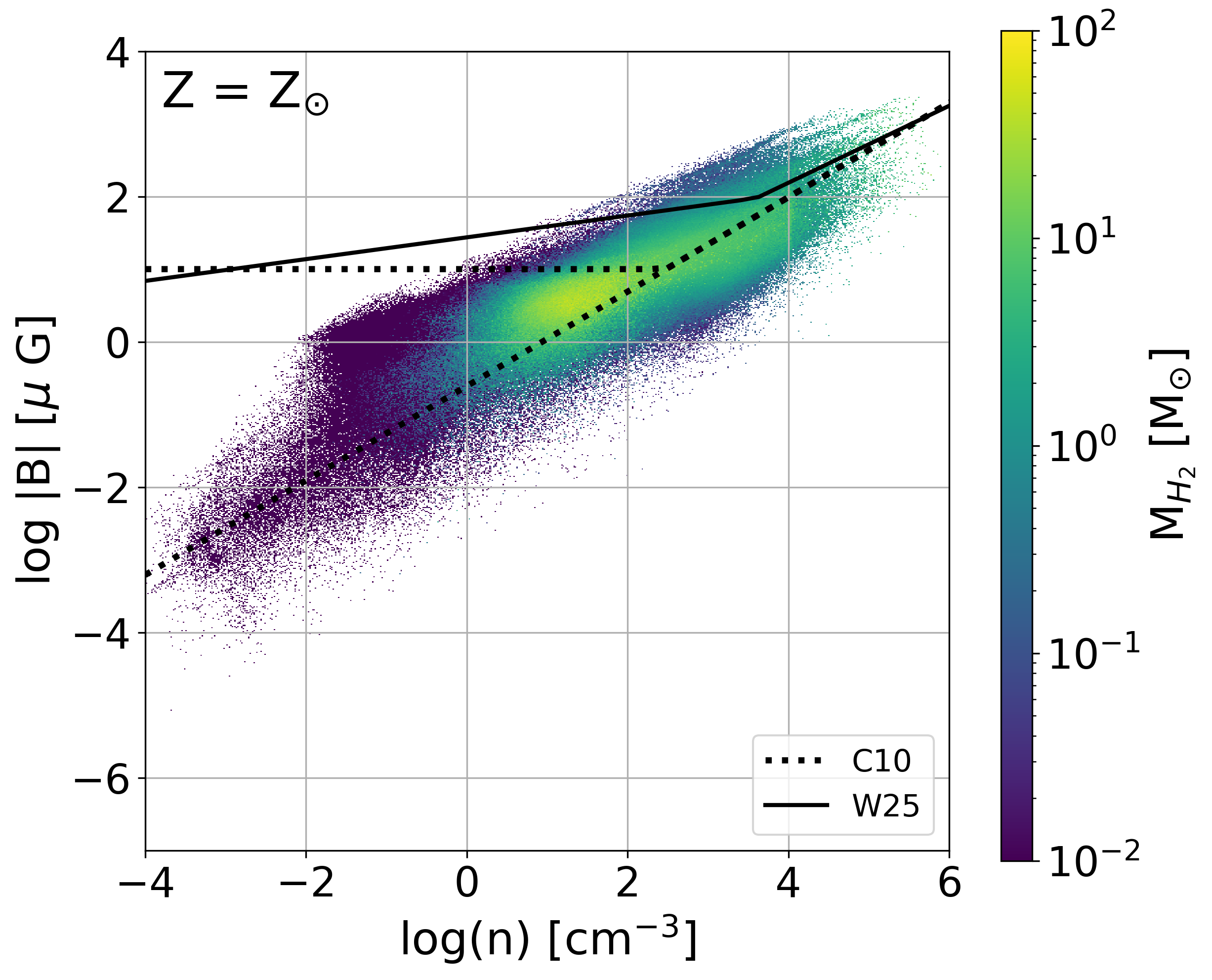}
    \caption{Mass-weighted absolute magnitude of the magnetic field vs the gas number density for at 1 Gyr for the solar metallicity model. The dotted black lines indicate the \protect \cite{Crutcher2010} relation, with and without a flattening at low density. The solid black line shows the slope derived in \citealt{Whitworth2025b}.}
    \label{fig:Bf_vs_n}
\end{figure}

Figure \ref{fig:Bf_vs_n} shows the absolute magnetic field strength, $|B|$, as a function of $n$ in the gas disc of our fiducial mode (where the disc is defined as $r<2.5$ kpc and $|z|<0.4$ kpc). Since we are investigating the effects of the strength of the magnetic field, it is important that these are in-line with observations. 
The grey dotted lines in Figure \ref{fig:Bf_vs_n} represents the observed B-$n$ relation, $\alpha = 0.66$ derived by \citet{Crutcher2010} from Zeeman observations and extrapolated to lower densities with and without the flattening originally proposed (note that the initial choice of a flat function at low densities was primarily due to lack of data). 

More recently, \citet{Whitworth2025b} performed a Bayesian meta-analysis on both observational data and numerical simulations showing that in the diffuse gas phase there is a correlation between the magnetic field strength and number density in opposition to the original results of \cite{Crutcher2010}. We plot the median fit derived from the local Zeeman data in \citet{Whitworth2025b} as a solid black line. An exact match between the numerical models and observed slope is not to be expected, as it depends on the specifics of field amplification in our dwarf galaxy given its internal turbulence. However, both the slope and normalisation of the relation are in reasonable agreement with observations and we are confident our field strengths are not over-estimates, see \citet{Whitworth2025b} for a detailed discussion on this topic.

\subsection{Energy analysis metrics}
\label{analysis_metrics}

\subsubsection{Averaging cells}
Before making any computations, one must choose between mass-weighting and volume-weighting the average gas properties, as the \textsc{Arepo} cells are irregular in size. The choice depends on which context we wish to analyse the simulation in. A mass-weighting scheme means that the quantity in which we are interested \textit{follows} the mass, as in a Lagrangian perspective. Conversely, a volume-weighted scheme essentially considers quantities at fixed packets of volume, as in an Eulerian perspective. We are more interested in where the mass is located in each gas phase and how it moves throughout the galaxy and molecular clouds within it; thus we adopt a mass-weighting scheme (a comparison of how choosing different weightings changes gas properties can be found in \citealt{Whitworth2025a}).
        
Each gas cell in the simulation contains a number of different chemical species that are tracked by our time-dependent model (see Section \ref{numerics}). However, while many species are available, some have very limited data due to their low abundance (e.g., more complex molecules such as HCO$^+$). Thus, we focus on four main species: HI, H$^+$, H$_2$, and CO as well as their combined behaviour. These species effectively probe the warm ionised medium (WIM), WNM, CNM, and molecular gas phases of the ISM. 

\subsubsection{Magnetic and thermal energy densities}
\label{energy_den1}
        
The magnitude of the various energy densities associated with the gas in the dwarf galaxy is perhaps the clearest indicator of which forces are dominant in the star-forming disc. For the magnetic and thermal energy densities, obtaining values for every gas cell is relatively easy. The thermal energy per unit mass ($E_{th}$), which is calculated within the simulation, is simply converted to thermal energy density by multiplying by the mass and dividing by the volume for each gas cell. The magnetic energy density is obtained via Equation~\ref{mag_den}, where $B$ is in cgs units as in \citetalias{Whitworth2023}.

\begin{equation}
    \label{mag_den}
    E_m = |\textbf{B}|^2 / 8\pi
\end{equation}

\subsubsection{Velocity dispersions}
\label{vel_dis}
    
Estimating the velocity dispersion is the greatest challenge in the analysis, as due to the turbulent nature of the ISM, the velocity dispersion will depend on the scale at which it is measured. From a numerical point of view, this poses difficulties as the spatial scale of an \textsc{Arepo} cell changes due to the adaptive nature of the mesh. Furthermore, the dispersion within a single cell is fundamentally a subgrid quantity, since it is mathematically impossible to calculate a dispersion from a single measurement. If we were to underestimate the turbulence, this would adversely affect our conclusions about magnetic field strength, and so we seek to make a conservative estimate that can serve as an upper limit of the local turbulent energy.
    
\begin{figure*}
        \centering
        \includegraphics[width=0.75\textwidth]{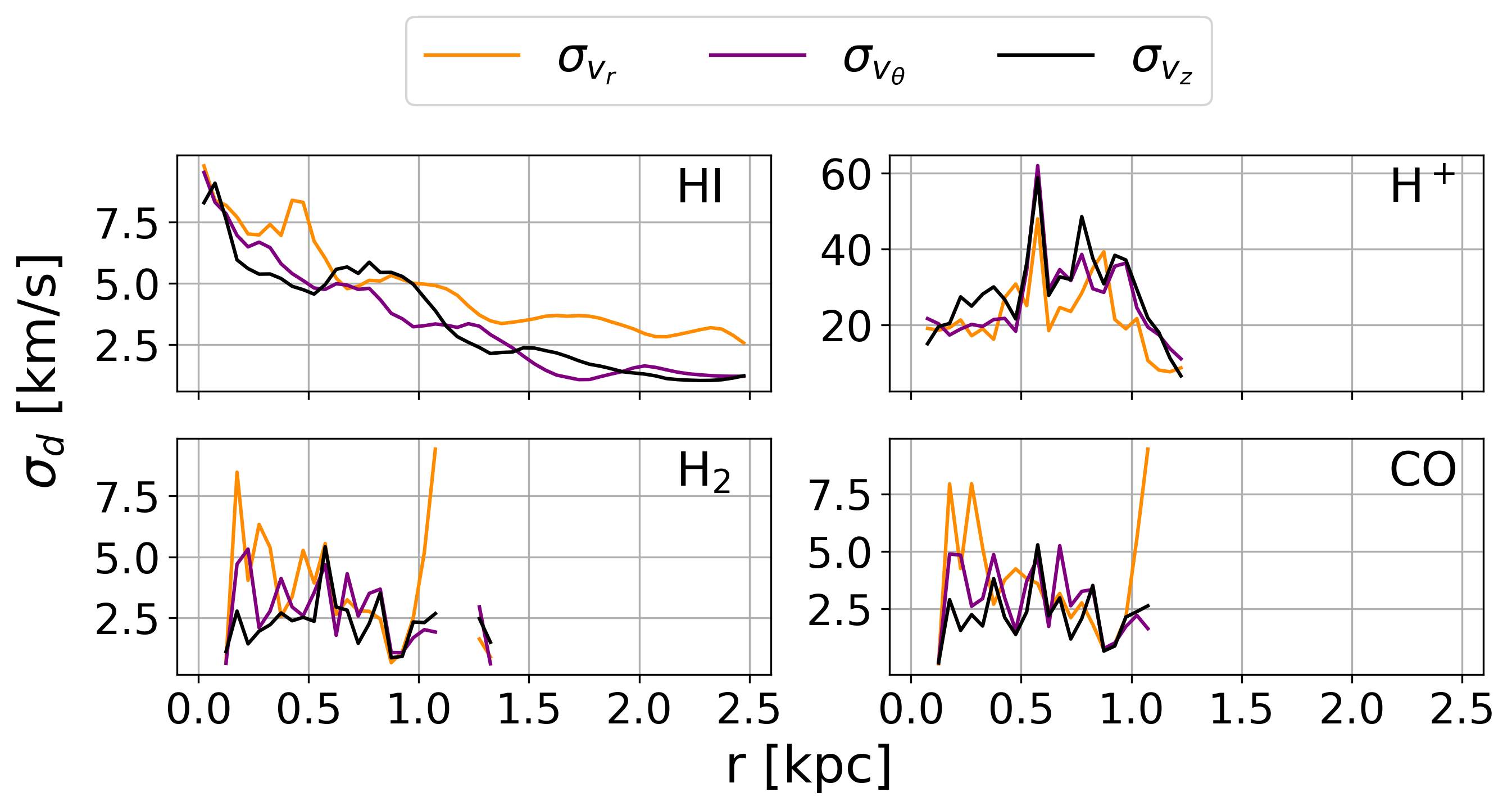}
        \caption{Radial profiles of the cylindrical components of the velocity dispersion for neutral, ionised and molecular hydrogen as well as CO.} 
        \label{fig:vel_dis}
    \end{figure*}

Firstly, we make a ballpark estimate of the turbulence within the galactic disc. To ensure we do not include galactic rotation in our estimate, we convert to cylindrical coordinates and then calculate the velocity dispersion in radial annuli. Under this coordinate system, the circular velocity will peak at the local rotation curve of the galaxy, and the dispersion will be reflective of turbulently driven departures from this. The velocity dispersions are calculated as in \citetalias{Whitworth2023} by dividing the disc into annuli bins from zero to 2.5 kpc with 50 pc widths. The mass-weighted velocity dispersion components are calculated for each bin via a weighted standard deviation of each velocity component, and the results are shown in Figure~\ref{fig:vel_dis}. We find that the magnitudes of the dispersions are similar for the cold and molecular gas species and have a range of 2-8 km~s$^{-1}$. The ionised gas speed is higher because of its presence in outflows, but it is a distinct gas phase from the atomic and molecular gas that is our focus, and so we neglect it for the rest of our analysis. 

\begin{figure}
        \centering
        \includegraphics[width=\columnwidth]{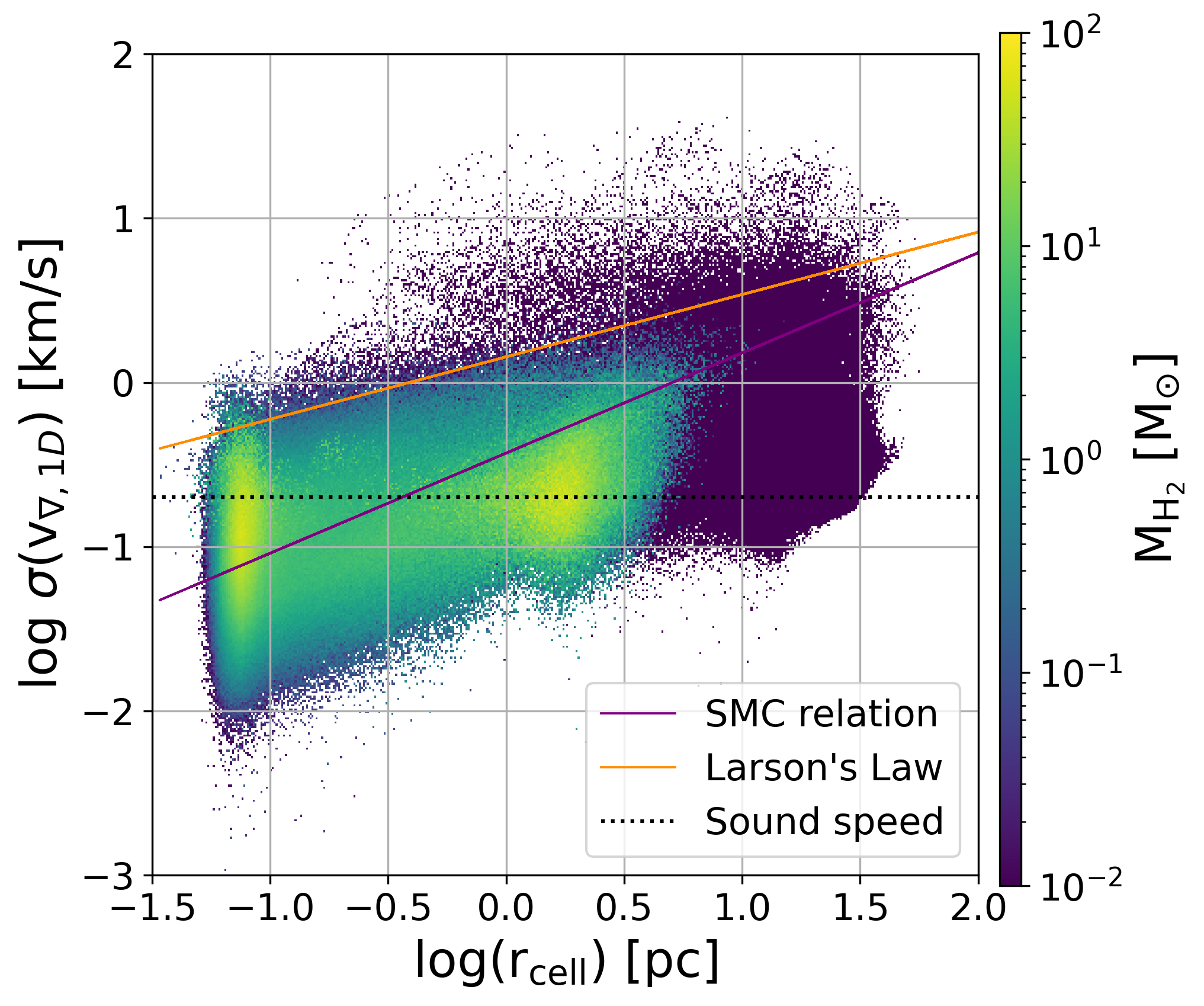}
        \caption{The velocity dispersion calculated for each cell using the gradients method as a function of cell size. The distribution is a weighted 2D histogram where the colour map indicates the total H$_2$ mass in each bin. Blue line: Larson's Law, derived from observations of Milky Way-like galaxies. Red line: The SMC relation from \citealt{Saldano2023}. Dotted line: The average sound speed within molecular clouds, c$_s$=0.2 km/s \citep[e.g.][]{MacLow&Klessen2004}}
        \label{fig:vgrad_vs_cellsize}
    \end{figure}

This is an upper limit for the velocity dispersion, as it is calculated over the entirety of each disc annulus. However, we want to compare the energy within single resolution elements of the models, as is done for the magnetic and thermal energy densities. One way of extracting this from the simulation is to use the velocity gradient across the cell that is computed in \textsc{Arepo} when calculating the fluxes to solve the Riemann problem. Each cell has nine velocity gradients associated with it, and taking the root-mean-square of these allows the calculation of the representative velocity gradient along the diameter of \textit{each} cell. To obtain an estimate of the velocity dispersion, we then multiply this gradient by the characteristic size of the cell such that 
    \begin{equation}
        \sigma(v_{\nabla,1D})= r_{\rm{cell}}\sqrt{\frac{1}{9} \sum_{n=1}^{9} \delta v_n^2 }
    \end{equation}
where $\delta v_n$ is the velocity gradient along each of the 9 directions. Figure \ref{fig:vgrad_vs_cellsize} shows the distribution of the velocity dispersion across each cell radius measured using the gradients method, where the cells are weighted by their internal molecular hydrogen mass. 

To validate this method, the measured cell velocity dispersions are compared to empirically observed size-velocity dispersion scaling relations. The most commonly used of these is Larson's relation~\citep[]{Larson1981}, which relates the observed velocity dispersion to the cloud size of the gas as shown below.
        \begin{equation}
            \label{eqn:larsens}
            \sigma (v_{1D}) [\rm{km/s}] = 1.10 L [\rm{pc}]^{0.38}
        \end{equation}        
If we take L to be the characteristic size of each cell in the simulation (i.e., $L=2 r_{\rm{cell}}$), we can then compute an empirical estimate of the unresolved velocity dispersion within each cell. 

This relation was derived for nearby clouds in the Milky Way, whereas our simulations are of dwarf galaxies. Consequently, we also compare to the velocity linewidth scaling relation of \cite{Saldano2023}, derived from a bootstrapping fit to observations of molecular clouds within the Small Magellanic Cloud (SMC).(As measured along a line of sight, this is a 1D dispersion.) This relation is shown in Equation~\ref{eqn:smc}, where we take R to be the radius of a cell. 
    \begin{multline}
        \label{eqn:smc}
        \log{(\sigma(v_{SMC})/\rm{km s}^{-1}) = \, -(0.43 \pm 0.18) +} \\
        {(0.61  \pm 0.25) \, \log(R/\rm{pc})}
    \end{multline}

In Figure \ref{fig:vgrad_vs_cellsize}, the measured cell velocity dispersions lie below the Larson scaling relation, which is consistent with expectations due to the lower supernova rate per unit area in the dwarf galaxy. However, the slope of the distribution is in good agreement. On the other hand, for the SMC relation, the magnitude of the velocity differences is in good agreement, but the scaling is different. \citet{Izquierdo2021} investigated the internal velocity dispersion of clouds in the Cloud Factory simulations of \citet{Smith2020} and found that the slope was invariant, but that the scaling depended on the amount of supernovae. It is unclear, therefore, if the change of slope seen in the SMC velocity dispersion relation is a real effect due to differences in the physics, or due to difficulty observing on small scales. Overall, our measured velocity difference seems a reasonable proxy for the internal velocity dispersion of a cell. It is lower than the vertical velocity dispersion of the disc from Figure \ref{fig:vel_dis}; it follows the slope of the velocity dispersion vs. size relation of \citet{Larson1981} from nearby molecular clouds where turbulent motions are well resolved; and it is consistent with the magnitudes of the velocity dispersions observed in the nearby SMC dwarf galaxy.

\subsubsection{Kinetic Energy}

Armed with a knowledge of the internal cell velocity dispersion, we next want to calculate the kinetic energy from turbulent motions. However, the adaptive mesh within \textsc{Arepo} introduces a potential bias to our method. Since the simulation resolution is dependent on density, the measurement scale over which we calculate the dispersion decreases as density increases (as shown in Figure \ref{fig:resolution_plot}). However, the above turbulent scaling laws ensure that the velocity dispersion will always be smaller as the measurement scale is decreased. Our goal is to determine where magnetic energy is most important in the energy balance of the ISM; therefore, it is important not to underestimate the other terms.

We therefore adopt a common scale of comparison. From our cell resolution plot, we see that there is a rapid decrease in cell size above densities of $10$ cm$^{-3}$ when the Jeans refinement criterion becomes active. The mean cell radius at this density is $1.15$ pc and so, we use $r_L=1.15$ pc as the minimum scale over which to measure turbulence. If we did not adopt this, then denser, more highly resolved cells would always have a lower velocity dispersion simply from a Larson-type turbulent scaling law. This density is where the thermal instability occurs and so ensures that we have adopted the correct length scale to test the hypothesis that magnetic fields are important in this regime. An added benefit is that the turbulence in the molecular gas in the simulation is measured across a common scale.  

Our detailed procedure for obtaining the effective velocity dispersion for the kinetic energy calculation is as follows. For cells smaller than $r_L$, we take the velocity dispersion to be the combination of the 1D velocity dispersion of the cell calculated using the gradient method $\sigma(v_{\nabla,1D})$, \textit{and} the mass-weighted velocity dispersion of the nearest neighbour cells within a radius $r_L$ of the cell centre $\sigma(v_{\rm{nn}})$. (We use an order of magnitude density contrast cut to ensure only coherent structure is included when calculating the velocity dispersion of the neighbouring cells $\sigma(v_{\rm{nn}})$ as discussed in the Appendix). For cells larger than $r_L$ we take the velocity dispersion to be the combination of the empirically derived SMC scaling relation at $r_L$,$\sigma(v_{\rm{SMC}})$, plus $\sigma(v_{\nabla,1D})$ calculated from the cell gradients. We chose the SMC law as it will have a supernova rate more representative of a dwarf galaxy and is a more sensible lower limit.

When comparing to observed scaling laws (which are necessarily along a single line-of-sight) it made sense to consider a 1D velocity dispersion. However, for calculating the kinetic energy, the 3D velocity dispersion is more relevant and so we convert the velocity dispersions to their 3D form by multiplying by $\sqrt{3}$.
Our procedure is summarised in Equation \ref{eq:velocity_dispersion}. We then use the empirically derived velocity dispersion from Equation \ref{eqn:smc} as a floor below which the velocity dispersion may not fall to further ensure our estimates are always upper limits.  In practice, this may lead to some double counting, but it guarantees our estimate for the importance of the magnetic field verses kinetic terms is always a conservative one.
    \begin{equation} \label{eq:velocity_dispersion}
    \sigma(v_{\rm{eff,3D}}{\rm}) = 
    \begin{cases}
    \sqrt{\sigma(v_{\nabla,3D})^2 + \sigma(v_{\rm{nn,3D}})^2 } & \text{if } r_{\rm{cell}} \le r_L \\
    \sqrt{\sigma(v_{\nabla,3D})^2 + \sigma(v_{\rm{SMC,3D}})^2 } & \text{if } r_{\rm{cell}} > r_L
    \end{cases}
    \end{equation}

\begin{figure}
    \centering
\includegraphics[width=\columnwidth]{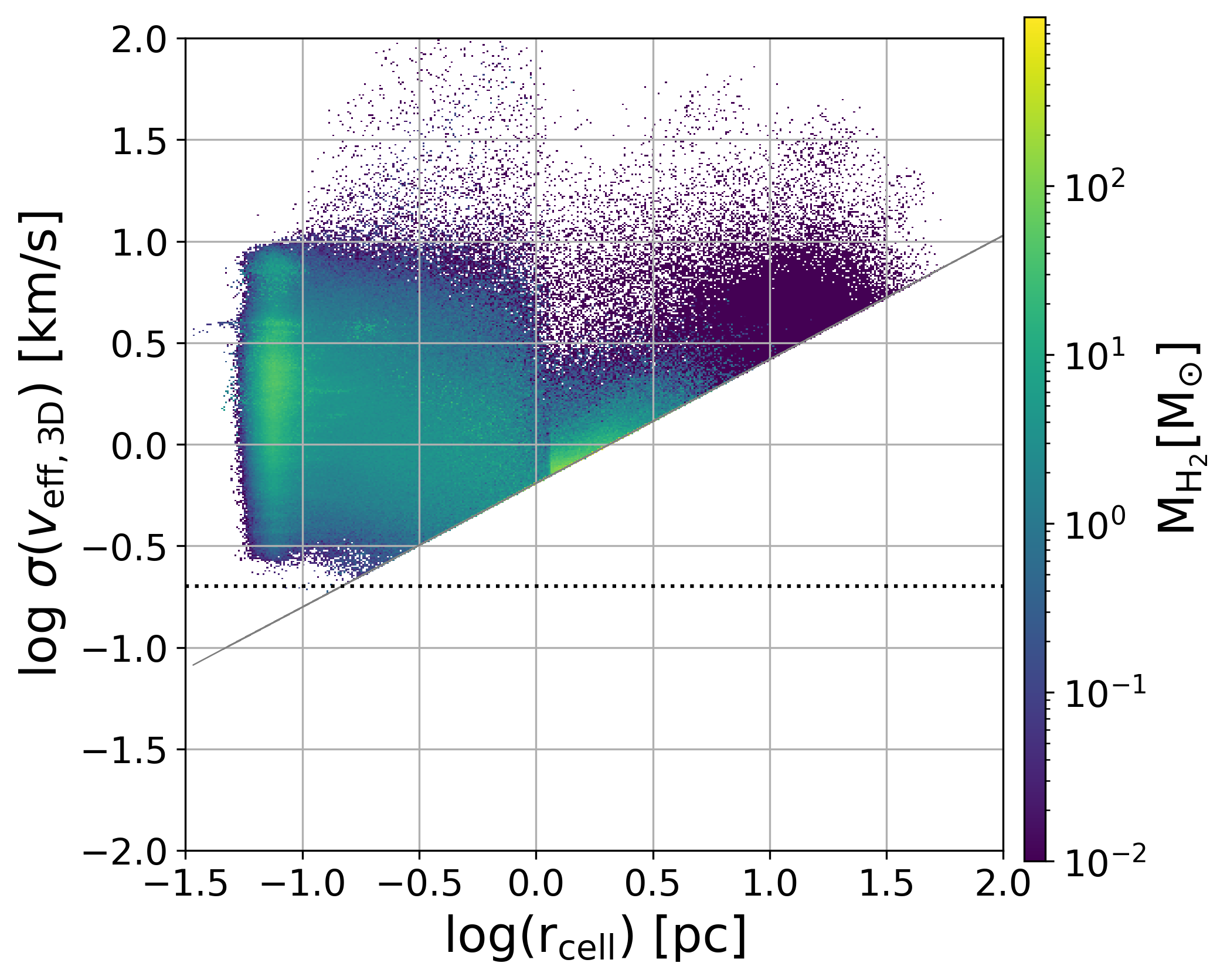}
    \caption{The distribution of effective 3D velocity dispersions used for calculating the kinetic energy of each cell. The dispersion is measured at a minimum scale of $r_L=1.15$ pc for all cells. The grey line shows the SMC scaling law (converted to 3D) which is used as a floor to ensure we never under-estimate the importance of turbulence compared to the magnetic field. }
    \label{fig:veff_vs_cellsize}
\end{figure}

Figure \ref{fig:veff_vs_cellsize} summarises the final effective velocity dispersion for each cell. Our distribution is both higher and flatter than the turbulent scaling relations at low cell sizes when measured at $r_L=1.15$~pc instead, as expected. The kinetic energy density can then be trivially calculated as:
    \begin{equation}
    \label{eqn:Ek}
        E_k = 1/2\: \rho \sigma({v_{\rm{eff,3D}}})^2
    \end{equation}
where $\rho$ is the mass density and $\sigma_{v_{\rm{eff,3D}}}$ is the total velocity dispersion \citep[e.g.,][]{Beck2015_Ek}.

\subsubsection{Mass-to-flux ratio}
Another useful quantity to consider is the mass-to-flux ratio, $M/\phi$. This indicates whether the gas within a projected area is collapsing due to gravity, $M/\phi \ge 1$, or is magnetically supported, $M/\phi < 1$. Thus, it can also be considered along with the kinetic, thermal, and magnetic energy densities to determine which of the four processes dominate. However, as will be seen in Section~\ref{comparing_ratios}, we require a unitless measure of the relative strength between self-gravity and magnetic forces. Thus, we use the mass-to-critical-mass ratio (adapted from \citealt{Myers2021}), where the critical mass, $M_B$, is the mass at which self-gravity dominates over magnetic forces, given in Equation~\ref{eqn:mass-to-crit-mass}.
\begin{equation}
    \rm M/M_B = \frac{2 \pi G^{1/2} m}{|B| V^{2/3}}
    \label{eqn:mass-to-crit-mass}
\end{equation}
Here m is the cell mass, |B| is the absolute magnetic field strength, and V is the cell volume, all in cgs units. This is computed for each gas cell in the simulation for comparison with the internal energy densities of the cells.

\section{Results}
\label{results}

\subsection{Energy Ratios}
\label{comparing_ratios}
To determine which of the four forces dominates within a cell, we calculate the ratios of the internal magnetic,$E_m$, and non-thermal turbulent kinetic energy, $E_k$ to the internal thermal energy, $E_{\rm{th}}$:
    \begin{equation}
        \beta_{\rm{th}} = E_{\rm{th}}/E_m
        \qquad
        \beta_{\rm{kin}} = E_k/E_m
        \qquad
        \label{eqn:ratios}
    \end{equation}
where $\beta_{\rm{th}}$ is the thermal $\beta$ and $\beta_{\rm{kin}}$ is an effective kinetic $\beta$. If $\beta_{\rm{kin}}$ > 1, this indicates that turbulent kinetic energy (hereafter just turbulence) dominates over magnetic forces, and if $\beta_{\rm{kin}}$ < 1, then magnetic forces dominate over turbulence (and similarly for $\beta_{\rm{th}}$). From these two ratios, we can determine which of the three energy densities is dominant. However, we must also consider the mass-to-critical mass ratio (Equation~\ref{eqn:mass-to-crit-mass}). This is especially true for H$_2$ and CO, since these gas species exist in molecular clouds that can contain gravitationally bound star-forming cores. 

HI and H$_2$ masses are simply calculated by multiplying the relevant chemical abundance by the gas cell mass, but as CO is merely a tracer of molecular gas, we apply an abundance condition $A_{\rm{CO}}>10^{-4}$ to identify all cells with significant amounts of CO, and then use their H$_2$ mass. This latter condition is taken from \cite{Smith2014}, and corresponds to the rough threshold where molecular gas transitions from CO-dark to CO-bright in a Milky-Way-disc-like environment. (Even in metal-poor dwarf galaxies, stars do form in molecular gas rather than atomic, though often it is CO-dark \citep{Whitworth2025a}).

    \begin{figure}
        \centering
        \includegraphics[width=\columnwidth]{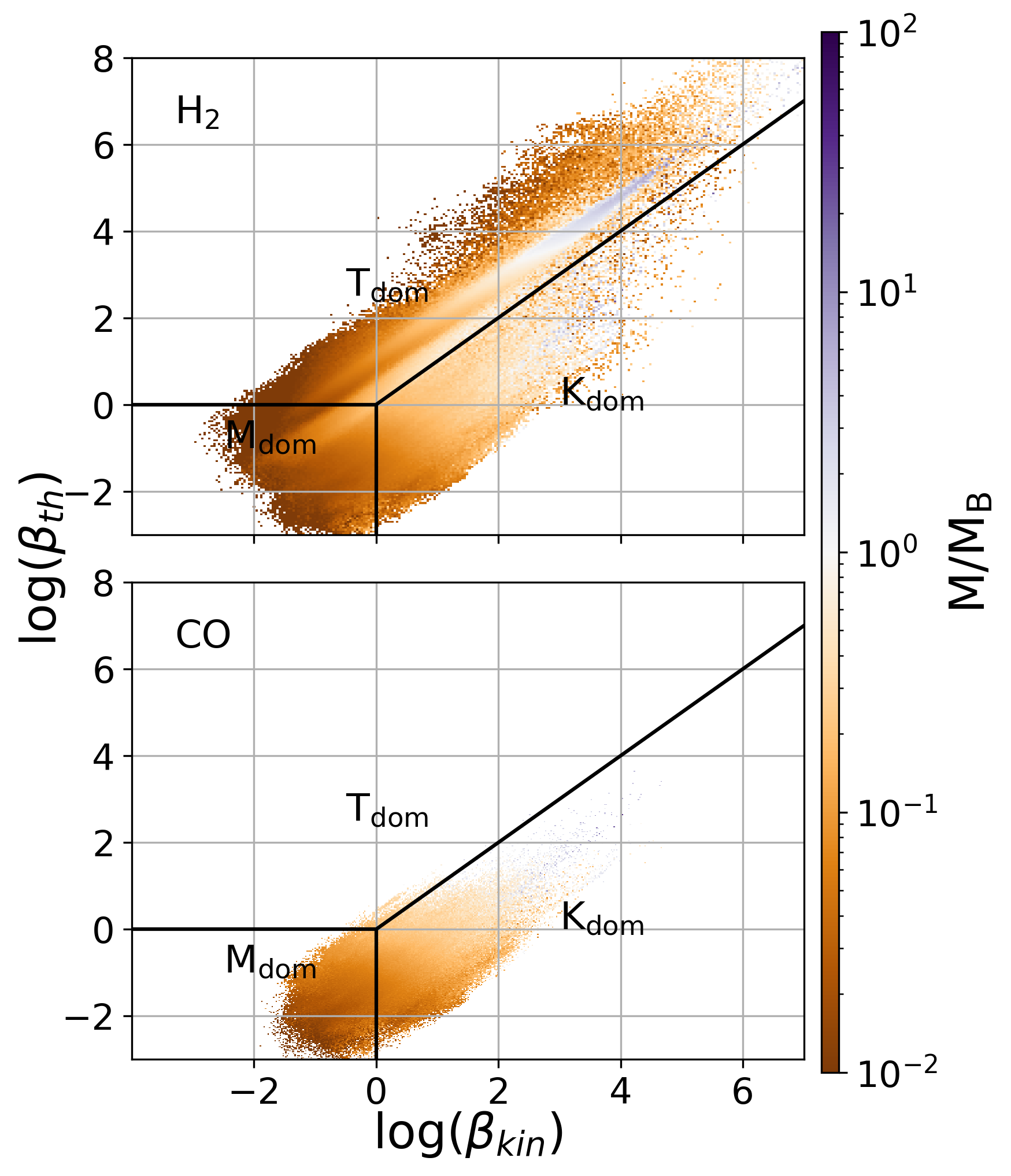}
        \caption{Energy ratios $\beta_{\rm{therm}}$ vs $\beta_{\rm{kin}}$ vs M/M$_B$ highlighting which forces dominate for all the molecular gas \textit{(top)} and only the molecular gas with substantial CO \textit{(bottom)}.}
        \label{fig:betaplot_H2}
    \end{figure}

After applying these cuts, we plot in Figure~\ref{fig:betaplot_H2} the distribution of $\beta_{\rm{kin}}$ vs. $\beta_{\rm{therm}}$. Each panel is split into three regions that indicate where magnetic forces dominate (M$_{\rm{dom}}$), thermal forces dominate (T$_{\rm{dom}}$), or turbulence dominates (K$_{\rm{dom}}$). The colour scale M/M$_{\rm{B}}$ shows in orange where magnetic forces are greater than those from gas self-gravity. Table~\ref{table:dominate_conditions} explicitly shows how these ratios can be used to form conditions to determine the dominant force in each cell. In Figure~\ref{fig:betaplot_H2}, most of the gas has M/M$_{\rm{B}}$ < 1, suggesting that there are very few gas pockets that are dominated by self-gravity. This is reasonable, since by definition, in these simulations, any gas that is dominated by self-gravity should form a sink particle and be removed from the gas phase. (Note that the purple sections that appear within the T$_{\rm{dom}}$ sections of Figure~\ref{fig:betaplot_H2} are unlikely to be self-gravitating due to the large values of $\beta_{\rm{therm}}$.) Naturally, this applies only to the cells' gas \textit{self-gravity} as large-scale gravitational fields could still be playing a role in the gas evolution, for example via large-scale collapse \citep{Smith2009b,Vazquez2019}, or tidal forces \citep{Ramirez2022}. The molecular H$_2$ is affected by all four forces, but the component traced by CO is dominated purely by magnetism, turbulence, and self-gravity.
    
\begin{table}
    \centering
    \begin{tabular}{lccr} 
    \hline
    Force & Condition \\
    \hline
    Turbulence  & $\beta_{\rm{th}}<\beta_{\rm{kin}}$ and $\beta_{\rm{kin}}>1$ and $\beta_{\rm{kin}}>M/M_B$ \\
    Thermal  & $\beta_{\rm{th}}>\beta_{\rm{kin}}$ and $\beta_{\rm{th}}>1$ and $\beta_{\rm{th}}>M/\phi$\\
    Magnetic & $\beta_{\rm{th}}<1$ and $\beta_{\rm{kin}}<1$ and $M/M_B$<1\\
    Self-Gravity  & $M/M_B>\beta_{\rm{th}}$ and $M/M_B>\beta_{\rm{kin}}$ and $M/M_B>1$\\
    \hline
    \end{tabular}
    \caption{Conditions under which each force dominates the gas phase as illustrated in Figure~\ref{fig:betaplot_H2}.}
    \label{table:dominate_conditions}
\end{table}

\subsection{Gas phases}
\label{phase_plots}
\begin{figure*}
    \centering
    \includegraphics[width=0.8\textwidth]{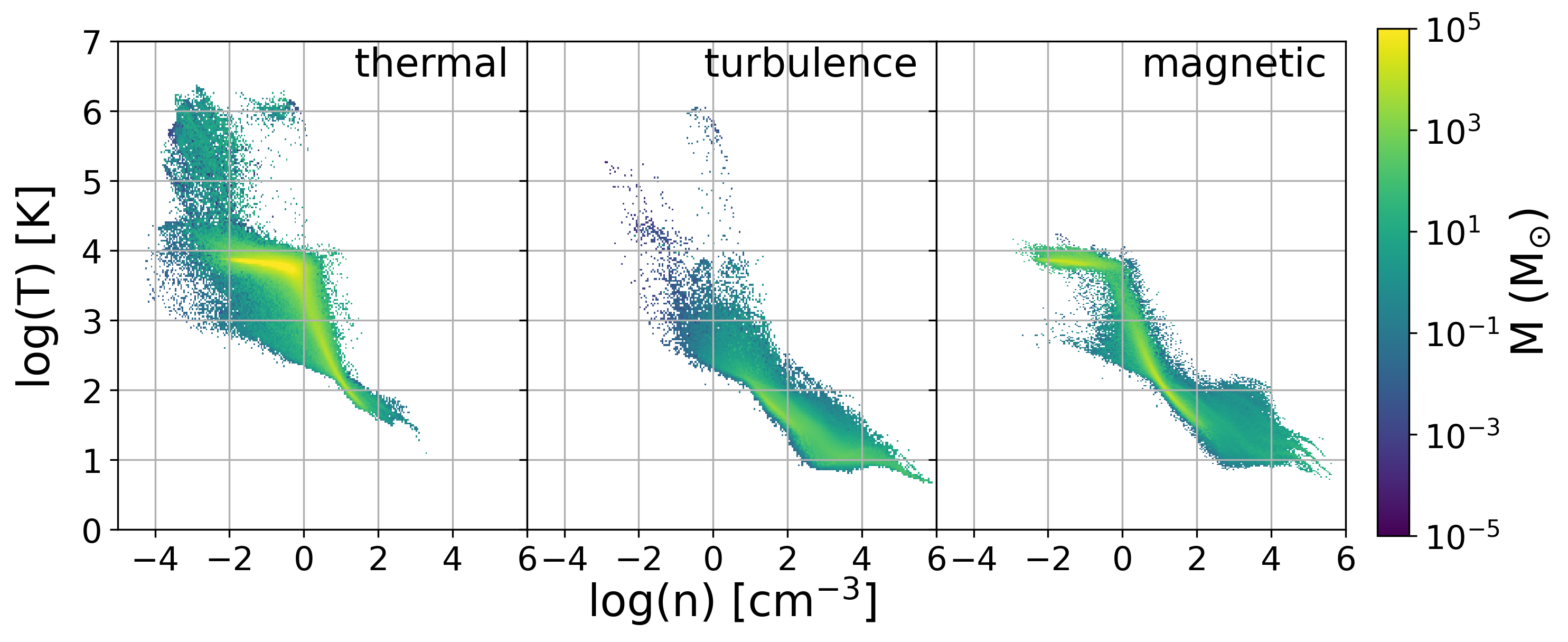} 
    \caption{Density-temperature phase plots highlighting where each force dominates. Self-gravity is not shown as the bulk of the self-gravitating mass is inside sink particles. The magnetically dominated gas lies at intermediate densities between thermally and turbulently dominated gas.}
    \label{fig:phase_plots}
\end{figure*}

Given the energetically dominant force for each gas cell in the simulation, we can use this to explore how these alter between the gas phases of the ISM. Figure~\ref{fig:phase_plots} shows 2D histograms of the distribution of total gas mass in density vs temperature space. Each panel shows only gas cells that are dominated by the labelled force. Self-gravity is not shown as the bulk of the self-gravitating mass is inside sink particles and so does not have a defined temperature and density. We emphasise that gravity may play a role at all scales due to the long-range nature of the force.

The thermally dominated regime in Figure~\ref{fig:phase_plots} peaks in the diffuse gas and although it encompasses a wide temperature range from $100-10^6$\,K, the most typical temperatures are $\sim10^4$~K as expected for the WNM. However, thermally dominated material is still found within the unstable regime between the WNM and CNM phases up to densities of $n\sim10^{2}$\,cm$^{-3}$. The turbulently dominated regime occurs at higher densities, up to $n\sim10^{5}$\,cm$^{-3}$, and at temperatures below $10^3$~K. It populates the phase space regime expected for the thermally unstable medium, the CNM, and molecular gas. The magnetically dominated material lies between the thermal and turbulence-dominated regimes. It is present in low density gas ($n\sim 10^{-2}$\,cm$^{-3}$) that will be part of the WNM up to densities commensurate with molecular gas ($n>100$\,cm$^{-3}$) and is prevalent in the thermally unstable regime in which warm gas is converted to cold gas clouds ($n\sim10$\,cm$^{-3}$).

\begin{figure}
    \centering
    \includegraphics[width=0.7
    \columnwidth]{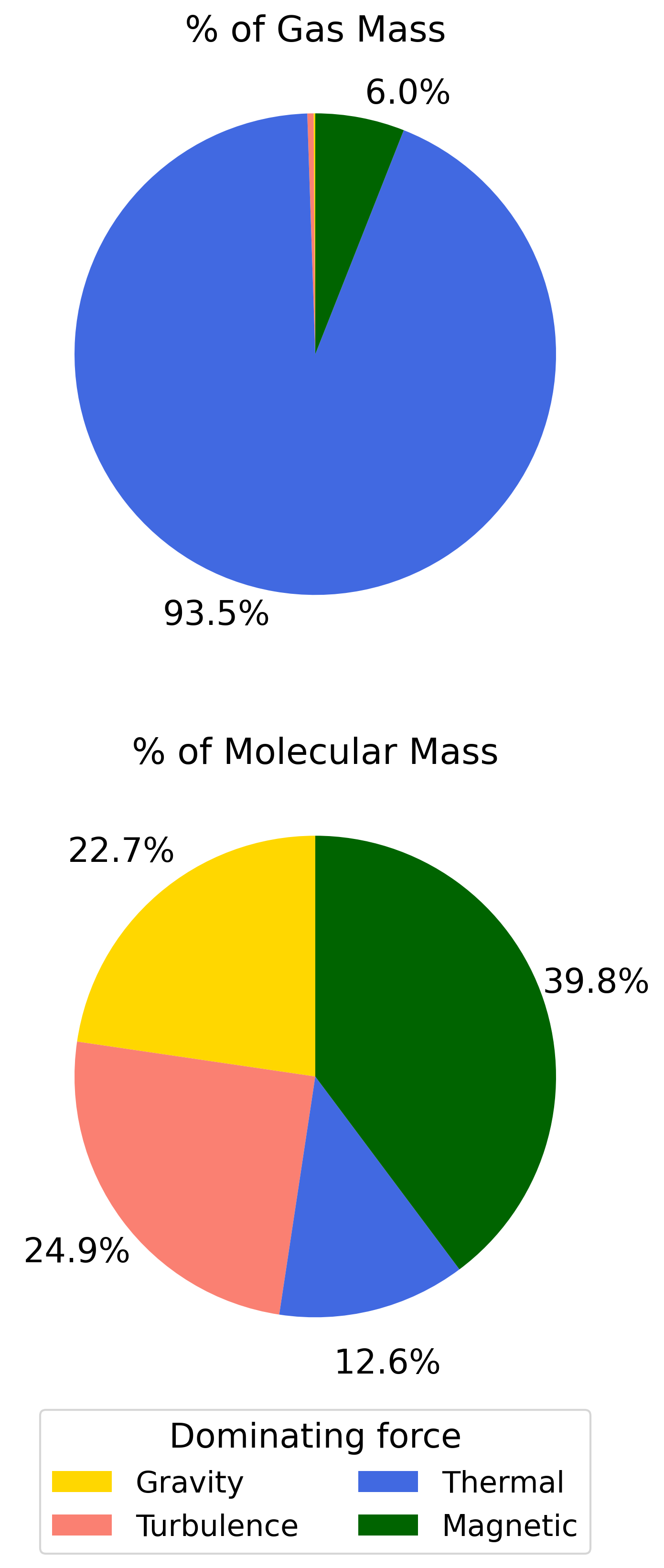}
    \caption{The fraction of the total, and molecular gas mass that is dominated by each force. The yellow segment refers only to self-gravity, and includes gas mass contained within young sinks. The total gas mass is mostly thermally dominated, as expected, however magnetic fields are the next most significant force. More of the molecular gas mass is dominated by magnetic energy than any other term.}
    \label{fig:pie_chart_combi}
\end{figure}

The total mass in each of the regimes is summarised as pie charts in Figure~\ref{fig:pie_chart_combi}
, which show the fraction of total gas and purely molecular gas mass respectively. The gas within young ($<2$~Myr) sink particles is counted as self-gravitating, as the criterion for sink particle formation within the simulation is that gas that is unambiguously bound and collapsing forms a sink. We assume that the gas mass inside these young sinks is still molecular. The total gas mass is mainly dominated by thermal energy ($93.5\%$), with magnetic energy ($6.0\%$) the next most important. Concentrating purely on the molecular gas where stars commonly form, there is a relatively even split between the regimes where each energy term dominates. The most significant energy term is magnetic energy ($39.8\%$), followed by the kinetic turbulent energy ($22.9\%$), and the mass dominated by gas self-gravity ($22.7\%$). Even in the molecular regime there is gas that is thermally dominated, however, comparing to Figure~\ref{fig:betaplot_H2} it is clear that this is not in the coldest H$_2$ that will contain visible CO (see \citealt{Glover2016} for an exploration of the temperature balance of the molecular phase). Likewise, much of the H$_2$ that is magnetically dominated may be CO-dark \citep{Smith2014, Whitworth2025a}.

\begin{figure}
    \centering
    \includegraphics[width=\columnwidth]{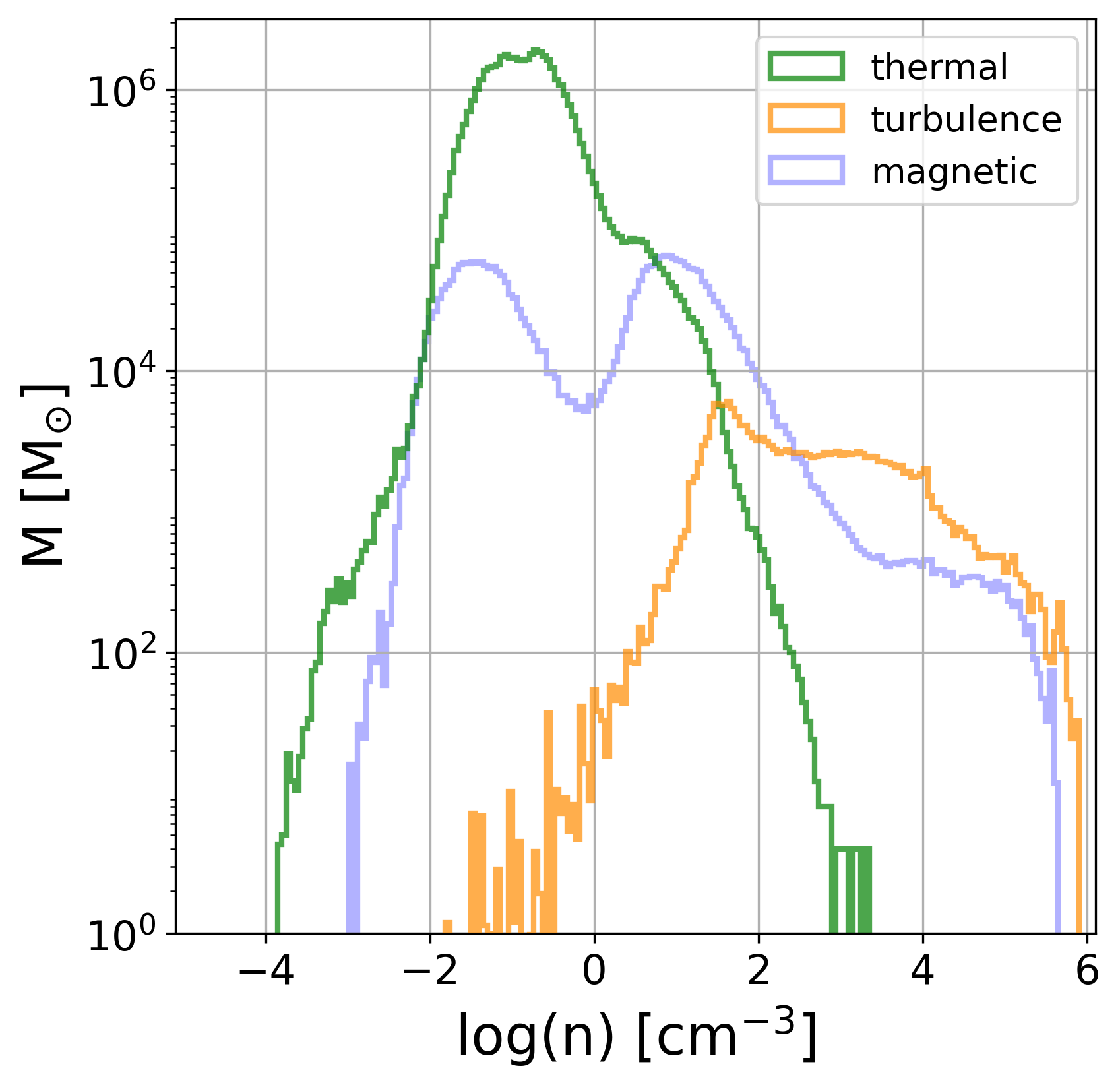}
    \caption{Distribution of binned gas mass vs. number density, highlighting that the different regimes peak at different densities. The magnetically dominated regime first dominates at number densities above $n\sim10$ cm$^{-3}$ typical of the thermally unstable regime where gas is is converted from warm to cold.}
    \label{fig:regime_hist}
\end{figure}

Figure \ref{fig:regime_hist} shows the distribution of the gas phase mass as a function of density. As in Figure \ref{fig:phase_plots}, thermal energies dominate at low densities, and kinetic at high densities. However, in between these regimes, magnetic energy dominates.


\begin{table}
    \centering
    \begin{tabular}{l|c|c|c|}
        \hline
        Regime & WNM  & Unstable & CNM \\
        \hline
        Thermal      & 96.9\% & 54.5\% & 27.7\% \\
        Magnetic     & 3.1 \% & 45.2\% & 66.1\% \\
        Turbulent    & 0.0\% & 0.3\% & 6.2\% \\
        \hline
    \end{tabular}
    \caption{The percentage composition of each of the atomic gas phases in terms of which energy dominates the gas mass in the phase.}
    \label{table:AtomicPhase_percentages}
\end{table}

Table \ref{table:AtomicPhase_percentages} shows the percentage of atomic gas mass dominated by each energy regime in the Warm, Cold, and Unstable gas phases. 
`WNM' refers to atomic mass with T $> 5000$\,K; `Unstable' refers to atomic gas with $5000\ge$ T $>200$\,K; and 'CNM' refers to atomic gas with T $\le200$\,K. 
We see that the WNM phase is dominated by thermal energy with a small influence from magnetic forces. The Unstable phase is roughly equally dominated by magnetic energy and thermal energy. The CNM is dominated by the magnetic energy with less thermal energy, as is expected due to its lower temperatures. For the turbulently dominated material, the bulk of the atomic gas mass is in the CNM. This emphasises the importance of magnetic fields in the transitional phase where warm gas is converted to cold gas in the simulation before molecular hydrogen forms.
        
\section{Discussion}



\subsection{Spatial distribution}
\label{cloud_study}

In Section~\ref{results} we focused purely on the density and temperature distribution of the interstellar medium within the dwarf Galaxy. We now investigate how this gas is located within the galaxy disc. 

\begin{figure*}
    \centering
    \includegraphics[width=0.9\textwidth,trim=0cm 0cm 0cm 1cm, clip]
    {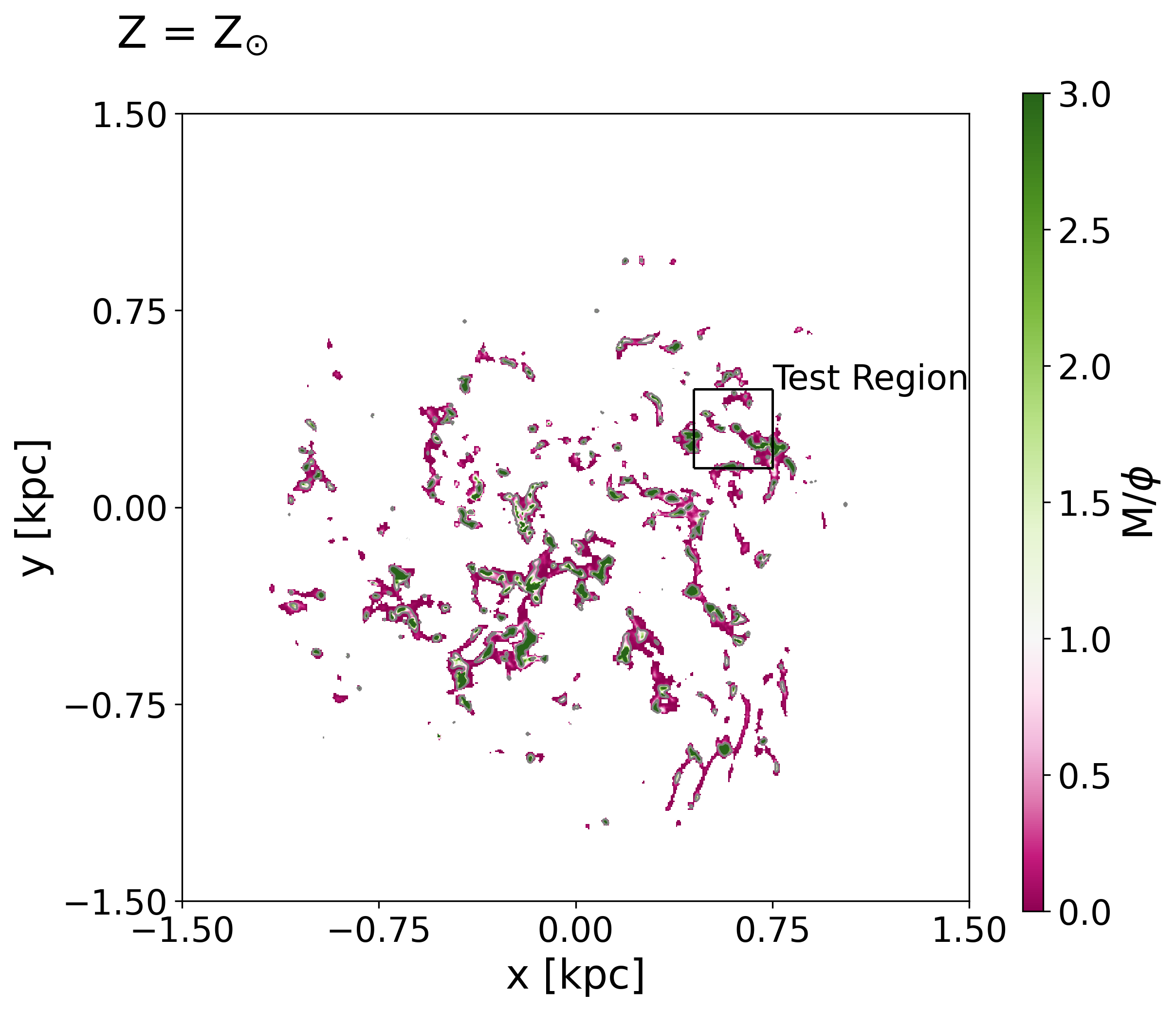}
    \caption{The mass-to-flux ratio in projection for the solar metallicity galaxy model. Note, the colour bar is anti-symmetric with white representing M/$\phi$ = 1. Gravity dominates (green) at the centre of the clouds but they are surrounded by sub-critical regions(purple) where magnetic forces dominate. A box shows the test region used for Figure~\ref{fig:cloud1}.}
    \label{fig:massflux}
\end{figure*}

Figure~\ref{fig:massflux} shows the mass-to-flux ratio calculated as a projection for the molecular gas in the galaxy disc. This is distinct from our earlier analysis, which was on a cell-by-cell basis in 3D, and instead refers to the balance of forces between gravity and magnetic fields across the full column of gas seen face-on in the disc. This gives a better indication of the gravitational forces compared to the magnetic forces at a larger scale compared to Section~\ref{comparing_ratios}, which only considers the self-gravity of the material within the cell. The mass-to-flux ratio $M/\phi$ is calculated as in \citet{Crutcher2004} using the relation $M/\phi =7.6\times 10^{-21} \rm N_{H_2}/|B|$ where N$_{\rm H_2}$ is the molecular column density in g\,cm$^{2}$ and |B| is the absolute magnitude of the field in $\mu$G. For clarity, we apply a density cut at N$_{\rm H_2}$ > 10$^{15}$\,cm$^{-2}$, to remove the most diffuse gas. In Figure~\ref{fig:massflux}, regions where gravity dominates over the magnetic field are shown in green and lie embedded at the centre of magnetically dominated regions. 

\begin{figure*}
    \centering
    \includegraphics[width=0.8\textwidth]{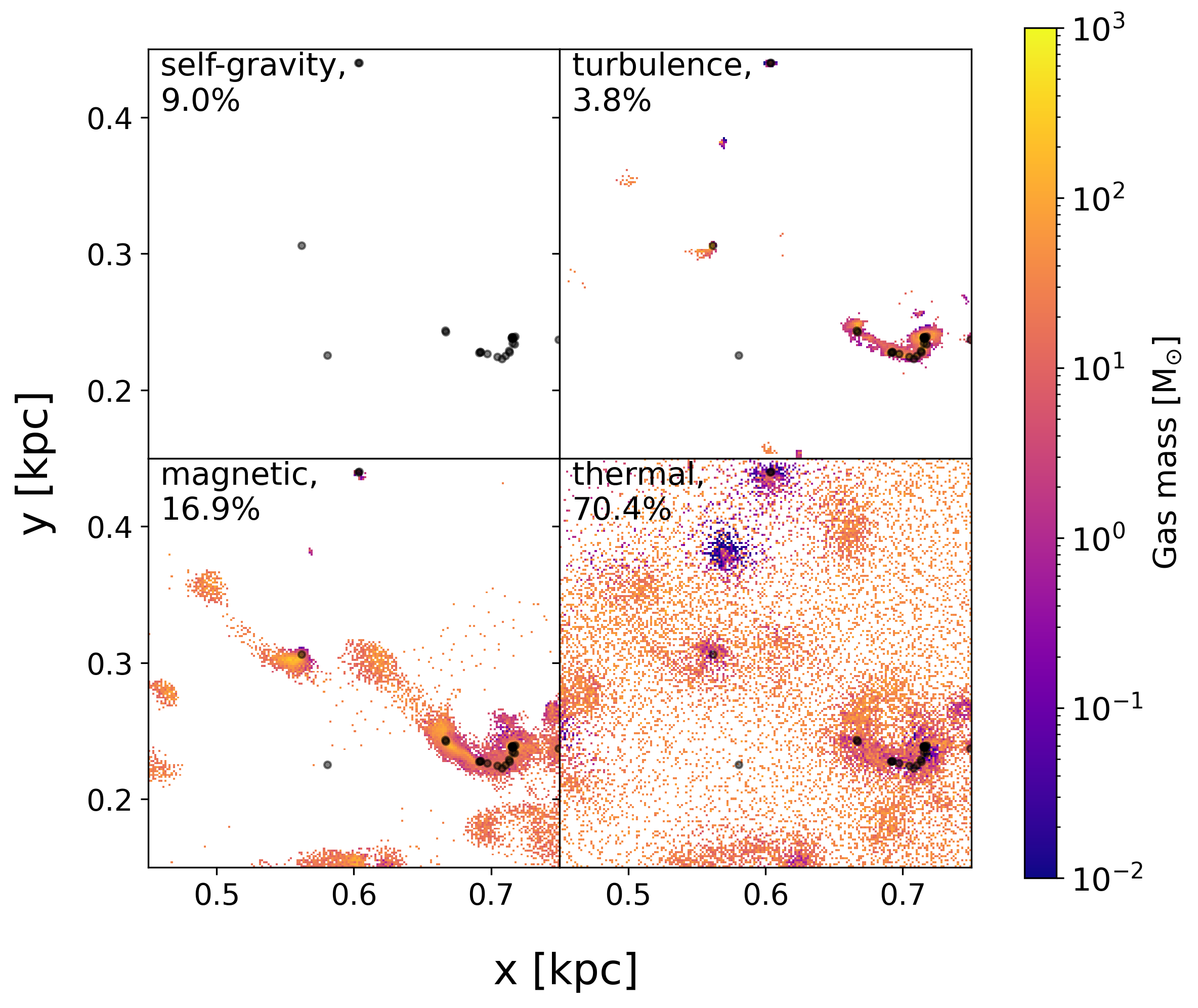}
    \caption{The location of and projected total gas mass in each energy 
    regime within the test region shown in Figure~\ref{fig:massflux}. The black dots indicate the location of sink particles, which are self-gravitating, and the percentages indicate the fraction of the total mass in the region in each energy dominated regime. Note, the mass percentages include mass within young sinks.}
    \label{fig:cloud1}
\end{figure*}

From Figure \ref{fig:massflux}, we chose a cubic test region (0.3 x 0.3 x 0.3 kpc) of gas that has clear supercritical structures (indicated in green) with surrounding subcritical structure (indicated in purple) so we may visually see how gas is distributed between these regimes. The location of gas cells dominated by each energetic term is shown in Figure~\ref{fig:cloud1}.

Most mass is in the thermal regime ($70.4\%$), which is distributed throughout the entire region. Next is the magnetic regime, containing $16.9\%$ of the mass. Not only does magnetically dominated gas surround the star forming-regions it also traces the filaments which lie between them (linking the large cluster in the bottom right up and to the left towards the smaller cores). After this is the turbulent regime, which dominates $3.8\%$ of the total gas mass in this region. This regime is only located close to the sink particles. This is the same behaviour as was found in~\cite{Li&Zhao2024}, where turbulence only dominates the gas after it has passed through a phase of being magnetically dominant. Lastly, is the self-gravity regime. There are only a handful of gas particles in this regime, which are not clearly visible in Figure~\ref{fig:cloud1} as they lie under the sink particle markers. The majority of the $9.0\%$ of mass in this regime comes from the young sink particles. (Note the total of these adds to $100.1\%$, but this is purely due to rounding.) Overall, when looking at Figure~\ref{fig:cloud1}, we see that, morphologically, the magnetic regime is an important phase in the gas evolution, linking the turbulent and thermal phases through filaments and likely permitting flows between them to some degree.

\subsection{Metallicity test}
\label{sec:metallicity}

\begin{figure}
    \centering
    \includegraphics[width=\linewidth]{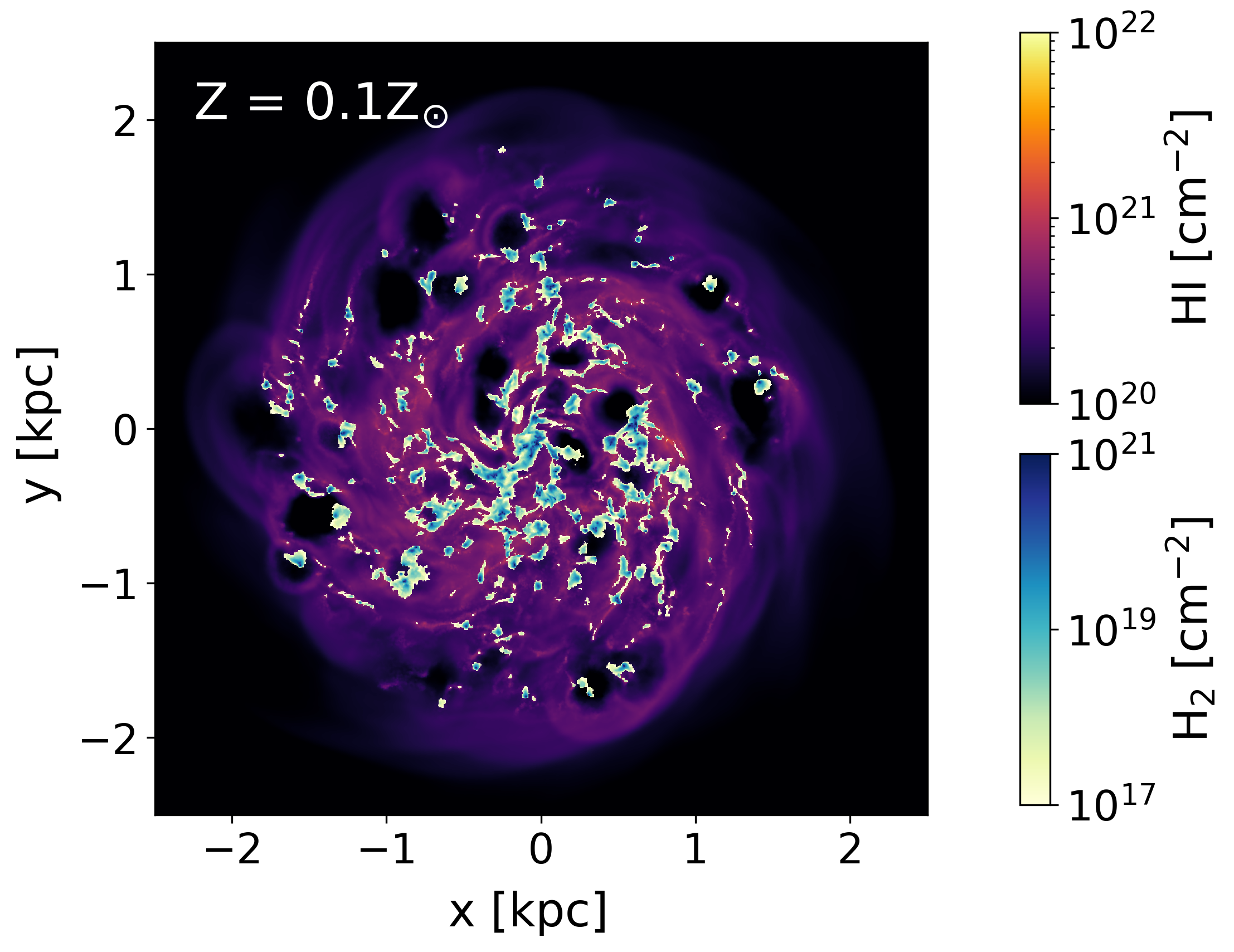}
    \caption{Projection of the HI and H$_2$ surface density for a low metallicity model taken from \citetalias{Whitworth2023}. The molecular gas is more widely distributed than for the solar case.} 
    \label{fig:SD_lowmetal}
\end{figure}

Our analysis in Section~\ref{results} used a single dwarf galaxy model of solar metallicity. To test the generalisability of these results, we revisit one of the original dwarf galaxy models from \citetalias{Whitworth2023}. This model has identical mass and initial conditions to our fiducial case but is at a metallicity of $0.1Z_\odot$, a $0.1\%$ dust-to-gas ratio and $10\%$ solar cosmic ionisation rate and interstellar radiation field. The surface density of the low-metallicity model can be seen in Figure~\ref{fig:SD_lowmetal}.

\begin{figure}
    \centering
    \includegraphics[width=\columnwidth, trim=0cm 0cm 0cm 1cm, clip]{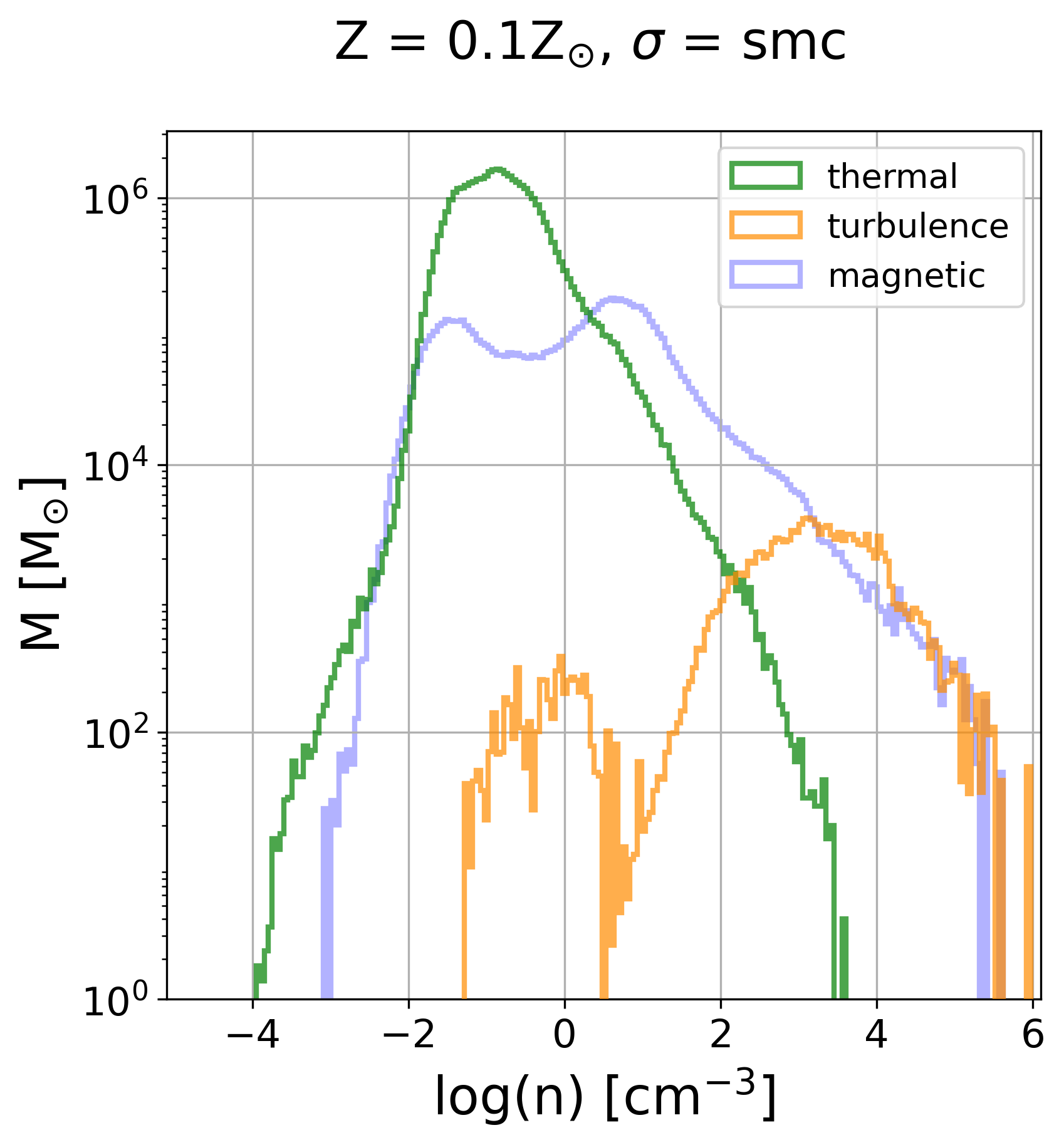}
    \caption{The estimated distribution of binned gas mass vs. number density from the low metallicity dwarf galaxy shown in Figure~\ref{fig:SD}. Given the more limited velocity dispersion information known for this simulation, the estimate is less precise, but the qualitative behaviour remains in agreement with Figure~\ref{fig:regime_hist}. }
    \label{fig:phase_plot_lowz}
\end{figure}

As this simulation was completed prior to the analysis we did here, the internal cell velocity gradients were not output when the model was run. Instead, we make a qualitative comparison using an effective velocity dispersion$\sigma(v_{\rm{eff,3D}})$ calculated purely from the combination of the velocity dispersion of cells within $r_L=1.15$~pc and the empirical SMC dispersion law such that $\sigma(v_{\rm{eff,3D}}{\rm}) = \sqrt{\sigma(v_{\rm{SMC,3D}})^2 + \sigma(v_{\rm{nn,3D}})^2 }$. The resulting mass histogram is shown in Figure~\ref{fig:phase_plot_lowz}. Given that this is a different approach to calculating the velocity dispersion compared to that used in Section~\ref{vel_dis}, the absolute amount of gas in the different phases will be a rough estimate. However, the qualitative behaviour is the same. Gas becomes energetically dominated by magnetic fields at $n\sim10$~cm$^{-3}$, intermediate between the thermal and kinetically-dominated regimes.

\subsection{A bottleneck for star formation?}
\label{bottleneck}
A consistent picture therefore emerges. During the conversion of warm gas to cold gas, magnetic fields will be energetically dominant. This is a particularly important transition during the star formation process. The cold neutral medium is the precursor to the formation of molecular clouds \citep[e.g.,][]{Liu2019,Syed2020} and simulations show that the cold neutral and atomic gas distribution overlap in space \citep{Smith2023} (although atomic and molecular gas filaments are often distinct objects that can have different orientations, as shown in \citealt{Soler2019}). Given the importance of magnetic fields in this regime, there will be non-negligible magnetic pressure forces that the gas must overcome in order to reach higher densities where the magnetic fields are less energetically important. Thus, magnetic fields act as a bottleneck during the star formation process at the galactic scale, and indeed during the entire baryon cycle.

The low-metallicity simulation discussed in Section \ref{sec:metallicity}, which was previously presented in \citetalias{Whitworth2023}, provides an illustrative example of how this bottleneck may change the composition of a galaxy disc. This paper found that including magnetic fields within the low-metallicity galaxy models increased the amount of H$_2$ and cold, dense gas contained within the molecular clouds compared to an identical purely hydrodynamical model, but did not change the star formation rate. \citetalias{Whitworth2023} conjectured that this is because the presence of the magnetic field slows the collapse of gas, slowing local star formation, meaning the molecular cloud assembly stage takes longer, resulting in a build-up of cold gas. Our energy analysis in this paper confirms that magnetic fields do dominate at densities associated with the formation of cold gas and so validates this argument. Our analysis also offers a natural explanation for polarisation observations that show that filaments of cold HI and magnetic fields are preferentially aligned \citep{Clark2014} but denser molecular clouds traced by CO are often not.

\subsection{Caveats}
Naturally, this analysis comes with several caveats. The simulations we use are single realisations of a dwarf galaxy model at two metallicities, and naturally the percentage of mass in each phase of the ISM will be to some extent model-dependent. Consequently, our percentage results should not be taken as some precise determination that can be applied in all environments. Rather, these are indicative fractions that show which physical forces are the most energetically important in different density and temperature regimes of the ISM. While the precise amount of mass in each phase may vary with galactic environment, the general finding that the magnetic field is most important during the transition between warm and cold gas will stand.

The greatest uncertainty we grappled with during the analysis was the scale at which turbulent energy should be measured, due to the multi-scale nature of this quantity. Our minimum measurement scale of $1.15$~pc was based upon the physical scale at which refinement of the underlying gas distribution was needed to continue to resolve the Jeans scale for gravitational collapse. This is commensurate with a typical size of star-forming clumps \citep{Bergin07} and gave sensible values when compared to the disc-averaged velocity dispersions in Figure~\ref{fig:vel_dis}. At all times, we tried to make our estimation of velocity dispersion a conservative one by combining both the internal cell velocity gradients, dispersion of nearby cells, and the empirically derived SMC scaling law from \citet{Saldano2023} as appropriate. When testing this for our fiducial case, we found our average was consistently higher than that suggested both by the Larson's scaling relation and the aforementioned SMC law. However, if a different approach had been taken this would have simply increased or decreased the fraction of the mass at high density that was turbulently dominated. The peak of the magnetically dominated regime would remain between the thermal and turbulent regime regardless of the choices made for the velocity dispersion calculation.

The galaxy models themselves also have uncertainties associated with them. Most important of these is probably the use of a background interstellar radiation field rather than distinct photoionising sources. For this reason, we have refrained from analysing the Hot Ionised Medium within this work. However, since the transition from WNM to CNM occurs well before photoionising stars are formed, this should not affect our results. 

\section{Conclusions}

We have used a dwarf galaxy simulation, of sufficiently high resolution to capture the formation of cold, dense gas down to sub-pc scales, to determine the energetic importance of the magnetic field with regard to other forces within the disc. 

The thermal, magnetic, turbulent kinetic, and self-gravitating potential energies of the gas enclosed in each cell of the simulation were compared to determine the density and temperature phase in which each dominated. To avoid being biased by the additional refinement needed within the simulation to capture the formation of the cold gas phase, we adopted a minimum measurement scale of $1.15$ pc when calculating the velocity dispersion needed for the turbulent kinetic energy. This is commensurate with the size scales of gas clumps within molecular clouds.

We find that the majority of gas in the galaxy is thermally dominated, as expected, but the next most important term is the magnetic energy. Within the cold neutral medium, and molecular hydrogen, magnetic energy is the most important term (albeit much of the molecular mass will be CO-dark). When computing the distributions in density and temperature phase space, there is a wide range of densities within which each force dominates over the others. However, the peaks of the distributions are well separated. The thermal energy is most important at number densities of $n\sim 1$~cm$^{-3}$; magnetic energies at number densities of $n\sim 10$~cm$^{-3}$; and turbulent energies above densities of $n\sim 100$~cm$^{-3}$. Gas self-gravity becomes dominant at the highest densities commensurate with the formation of collapsing cores (represented by sink particles in our simulations). We stress that our analysis only applies to the gravitational potential energy of the mass within a gas cell, and that larger-scale gravitational collapse motions could always be present. The distribution of the gas densities for cells that are magnetically dominated overlaps with the density range over which the thermal instability that converts the warm neutral medium to the cold neutral medium in the interstellar medium occurs. We therefore suggest that overcoming magnetic forces is an important bottleneck which must be overcome when forming the cold, dense gas clouds needed for star formation.

\section*{Acknowledgements}

RM acknowledges all of the support received from fellow students, in particular Laura Just-Fung, during his MPhys project at the University of St. Andrews. DJW acknowledges support from the Programa de Becas posdoctorales of the Direcci \'{o}n General de Asuntos del Personal Acad\'{e}mico of the Universidad Nacional Aut\'{o}noma de M\'{e}xico (DGAPA,UNAM,Mexico).

This work used the DiRAC@Durham facility managed by the Institute for Computational Cosmology on behalf of the STFC DiRAC HPC Facility (www.dirac.ac.uk). The equipment was funded by BEIS capital funding via STFC capital grants ST/P002293/1, ST/R002371/1 and ST/S002502/1, Durham University and STFC operations grant ST/R000832/1. DiRAC is part of the National e-Infrastructure.

\section*{Data Availability}

Projection grids of the fiducial dwarf galaxy are available at https://zenodo.org/records/17482943. Further data (and assistance in using it) is available upon request from either RJS or DJW.
 



\bibliographystyle{mnras}
\bibliography{References} 

\begin{thebibliography}{}
\makeatletter
\relax
\def\mn@urlcharsother{\let\do\@makeother \do\$\do\&\do\#\do\^\do\_\do\%\do\~}
\def\mn@doi{\begingroup\mn@urlcharsother \@ifnextchar [ {\mn@doi@} {\mn@doi@[]}}
\def\mn@doi@[#1]#2{\def\@tempa{#1}\ifx\@tempa\@empty \href {http://dx.doi.org/#2} {doi:#2}\else \href {http://dx.doi.org/#2} {#1}\fi \endgroup}
\def\mn@eprint#1#2{\mn@eprint@#1:#2::\@nil}
\def\mn@eprint@arXiv#1{\href {http://arxiv.org/abs/#1} {{\tt arXiv:#1}}}
\def\mn@eprint@dblp#1{\href {http://dblp.uni-trier.de/rec/bibtex/#1.xml} {dblp:#1}}
\def\mn@eprint@#1:#2:#3:#4\@nil{\def\@tempa {#1}\def\@tempb {#2}\def\@tempc {#3}\ifx \@tempc \@empty \let \@tempc \@tempb \let \@tempb \@tempa \fi \ifx \@tempb \@empty \def\@tempb {arXiv}\fi \@ifundefined {mn@eprint@\@tempb}{\@tempb:\@tempc}{\expandafter \expandafter \csname mn@eprint@\@tempb\endcsname \expandafter{\@tempc}}}

\bibitem[\protect\citeauthoryear{{Abe}, {Inoue}, {Inutsuka}  \& {Matsumoto}}{{Abe} et~al.}{2021}]{Abe2021}
{Abe} D.,  {Inoue} T.,  {Inutsuka} S.-i.,   {Matsumoto} T.,  2021, \mn@doi [\apj] {10.3847/1538-4357/ac07a1}, \href {https://ui.adsabs.harvard.edu/abs/2021ApJ...916...83A} {916, 83}

\bibitem[\protect\citeauthoryear{{Angl{\'e}s-Alc{\'a}zar}, {Faucher-Gigu{\`e}re}, {Kere{\v{s}}}, {Hopkins}, {Quataert}  \& {Murray}}{{Angl{\'e}s-Alc{\'a}zar} et~al.}{2017}]{Angles2017}
{Angl{\'e}s-Alc{\'a}zar} D.,  {Faucher-Gigu{\`e}re} C.-A.,  {Kere{\v{s}}} D.,  {Hopkins} P.~F.,  {Quataert} E.,   {Murray} N.,  2017, \mn@doi [\mnras] {10.1093/mnras/stx1517}, \href {https://ui.adsabs.harvard.edu/abs/2017MNRAS.470.4698A} {470, 4698}

\bibitem[\protect\citeauthoryear{{Bate}, {Bonnell}  \& {Price}}{{Bate} et~al.}{1995}]{Bate1995}
{Bate} M.~R.,  {Bonnell} I.~A.,   {Price} N.~M.,  1995, \mn@doi [\mnras] {10.1093/mnras/277.2.362}, \href {https://ui.adsabs.harvard.edu/abs/1995MNRAS.277..362B} {277, 362}

\bibitem[\protect\citeauthoryear{{Beck}}{{Beck}}{2015}]{Beck2015_Ek}
{Beck} R.,  2015, \mn@doi [\aapr] {10.1007/s00159-015-0084-4}, \href {https://ui.adsabs.harvard.edu/abs/2015A&ARv..24....4B} {24, 4}

\bibitem[\protect\citeauthoryear{Beck, Brandenburg, Moss, Shukurov  \& Sokoloff}{Beck et~al.}{1996}]{Beck1996}
Beck R.,  Brandenburg A.,  Moss D.,  Shukurov A.,   Sokoloff D.,  1996, \mn@doi [Annual Review of Astronomy and Astrophysics] {10.1146/annurev.astro.34.1.155}, 34, 155

\bibitem[\protect\citeauthoryear{{Bergin} \& {Tafalla}}{{Bergin} \& {Tafalla}}{2007a}]{Bergin2007}
{Bergin} E.~A.,  {Tafalla} M.,  2007a, \araa, 45, 339

\bibitem[\protect\citeauthoryear{{Bergin} \& {Tafalla}}{{Bergin} \& {Tafalla}}{2007b}]{Bergin07}
{Bergin} E.~A.,  {Tafalla} M.,  2007b, \mn@doi [\araa] {10.1146/annurev.astro.45.071206.100404}, \href {https://ui.adsabs.harvard.edu/abs/2007ARA&A..45..339B} {45, 339}

\bibitem[\protect\citeauthoryear{{Chevance}, {Krumholz}, {McLeod}, {Ostriker}, {Rosolowsky}  \& {Sternberg}}{{Chevance} et~al.}{2023}]{Chevance2023}
{Chevance} M.,  {Krumholz} M.~R.,  {McLeod} A.~F.,  {Ostriker} E.~C.,  {Rosolowsky} E.~W.,   {Sternberg} A.,  2023, in {Inutsuka} S.,  {Aikawa} Y.,  {Muto} T.,  {Tomida} K.,   {Tamura} M.,  eds,  Astronomical Society of the Pacific Conference Series Vol. 534, Protostars and Planets VII. p.~1 (\mn@eprint {arXiv} {2203.09570}), \mn@doi{10.48550/arXiv.2203.09570}

\bibitem[\protect\citeauthoryear{{Clark} \& {Hensley}}{{Clark} \& {Hensley}}{2019}]{Clark2019}
{Clark} S.~E.,  {Hensley} B.~S.,  2019, \mn@doi [\apj] {10.3847/1538-4357/ab5803}, \href {https://ui.adsabs.harvard.edu/abs/2019ApJ...887..136C} {887, 136}

\bibitem[\protect\citeauthoryear{{Clark}, {Glover}  \& {Klessen}}{{Clark} et~al.}{2012}]{Clark2012}
{Clark} P.~C.,  {Glover} S. C.~O.,   {Klessen} R.~S.,  2012, \mn@doi [\mnras] {10.1111/j.1365-2966.2011.20087.x}, \href {https://ui.adsabs.harvard.edu/abs/2012MNRAS.420..745C} {420, 745}

\bibitem[\protect\citeauthoryear{{Clark}, {Peek}  \& {Putman}}{{Clark} et~al.}{2014}]{Clark2014}
{Clark} S.~E.,  {Peek} J.~E.~G.,   {Putman} M.~E.,  2014, \mn@doi [\apj] {10.1088/0004-637X/789/1/82}, \href {https://ui.adsabs.harvard.edu/abs/2014ApJ...789...82C} {789, 82}

\bibitem[\protect\citeauthoryear{{Crutcher}}{{Crutcher}}{2004}]{Crutcher2004}
{Crutcher} R.~M.,  2004, \mn@doi [\apss] {10.1023/B:ASTR.0000045021.42255.95}, \href {https://ui.adsabs.harvard.edu/abs/2004Ap&SS.292..225C} {292, 225}

\bibitem[\protect\citeauthoryear{{Crutcher}}{{Crutcher}}{2012}]{Crutcher2012}
{Crutcher} R.~M.,  2012, \mn@doi [\araa] {10.1146/annurev-astro-081811-125514}, \href {https://ui.adsabs.harvard.edu/abs/2012ARA&A..50...29C} {50, 29}

\bibitem[\protect\citeauthoryear{{Crutcher}, {Wandelt}, {Heiles}, {Falgarone}  \& {Troland}}{{Crutcher} et~al.}{2010}]{Crutcher2010}
{Crutcher} R.~M.,  {Wandelt} B.,  {Heiles} C.,  {Falgarone} E.,   {Troland} T.~H.,  2010, \mn@doi [\apj] {10.1088/0004-637X/725/1/466}, \href {https://ui.adsabs.harvard.edu/abs/2010ApJ...725..466C} {725, 466}

\bibitem[\protect\citeauthoryear{{Dale}}{{Dale}}{2015}]{Dale2015}
{Dale} J.~E.,  2015, \mn@doi [\nar] {10.1016/j.newar.2015.06.001}, \href {https://ui.adsabs.harvard.edu/abs/2015NewAR..68....1D} {68, 1}

\bibitem[\protect\citeauthoryear{{Draine}}{{Draine}}{1978}]{Draine1978}
{Draine} B.~T.,  1978, \mn@doi [\apjs] {10.1086/190513}, \href {https://ui.adsabs.harvard.edu/abs/1978ApJS...36..595D} {36, 595}

\bibitem[\protect\citeauthoryear{{Draine}}{{Draine}}{2011}]{Draine2011_book}
{Draine} B.~T.,  2011, Physics of the Interstellar and Intergalactic Medium..
Princeton Series in Astrophysics, Princeton University Press, \url {https://search.ebscohost.com/login.aspx?direct=true&amp;AuthType=sso&amp;db=nlebk&amp;AN=1922543&amp;site=ehost-live&amp;authtype=sso&amp;custid=s3011414}

\bibitem[\protect\citeauthoryear{{Field}, {Goldsmith}  \& {Habing}}{{Field} et~al.}{1969}]{Field1969}
{Field} G.~B.,  {Goldsmith} D.~W.,   {Habing} H.~J.,  1969, \mn@doi [\apjl] {10.1086/180324}, \href {https://ui.adsabs.harvard.edu/abs/1969ApJ...155L.149F} {155, L149}

\bibitem[\protect\citeauthoryear{{Ganguly}, {Walch}, {Clarke}  \& {Seifried}}{{Ganguly} et~al.}{2024}]{Ganguly2024}
{Ganguly} S.,  {Walch} S.,  {Clarke} S.~D.,   {Seifried} D.,  2024, \mn@doi [\mnras] {10.1093/mnras/stae032}, \href {https://ui.adsabs.harvard.edu/abs/2024MNRAS.528.3630G} {528, 3630}

\bibitem[\protect\citeauthoryear{{Girichidis}, {Seifried}, {Naab}, {Peters}, {Walch}, {W{\"u}nsch}, {Glover}  \& {Klessen}}{{Girichidis} et~al.}{2018}]{Girichidis2018}
{Girichidis} P.,  {Seifried} D.,  {Naab} T.,  {Peters} T.,  {Walch} S.,  {W{\"u}nsch} R.,  {Glover} S. C.~O.,   {Klessen} R.~S.,  2018, \mn@doi [\mnras] {10.1093/mnras/sty2016}, \href {https://ui.adsabs.harvard.edu/abs/2018MNRAS.480.3511G} {480, 3511}

\bibitem[\protect\citeauthoryear{{Glover} \& {Smith}}{{Glover} \& {Smith}}{2016}]{Glover2016}
{Glover} S. C.~O.,  {Smith} R.~J.,  2016, \mnras, 462, 3011

\bibitem[\protect\citeauthoryear{{Gong}, {Ostriker}  \& {Wolfire}}{{Gong} et~al.}{2017}]{Gong2017}
{Gong} M.,  {Ostriker} E.~C.,   {Wolfire} M.~G.,  2017, \mn@doi [\apj] {10.3847/1538-4357/aa7561}, \href {https://ui.adsabs.harvard.edu/abs/2017ApJ...843...38G} {843, 38}

\bibitem[\protect\citeauthoryear{{Greif}, {Springel}, {White}, {Glover}, {Clark}, {Smith}, {Klessen}  \& {Bromm}}{{Greif} et~al.}{2011}]{Greif11}
{Greif} T.~H.,  {Springel} V.,  {White} S. D.~M.,  {Glover} S. C.~O.,  {Clark} P.~C.,  {Smith} R.~J.,  {Klessen} R.~S.,   {Bromm} V.,  2011, \mn@doi [\apj] {10.1088/0004-637X/737/2/75}, \href {https://ui.adsabs.harvard.edu/abs/2011ApJ...737...75G} {737, 75}

\bibitem[\protect\citeauthoryear{{Grisdale}, {Agertz}, {Romeo}, {Renaud}  \& {Read}}{{Grisdale} et~al.}{2017}]{Grisdale2017}
{Grisdale} K.,  {Agertz} O.,  {Romeo} A.~B.,  {Renaud} F.,   {Read} J.~I.,  2017, \mnras, 466, 1093

\bibitem[\protect\citeauthoryear{{Hennebelle}}{{Hennebelle}}{2013}]{Hennebelle2013a}
{Hennebelle} P.,  2013, \aap, 556, A153

\bibitem[\protect\citeauthoryear{{Hennebelle} \& {Inutsuka}}{{Hennebelle} \& {Inutsuka}}{2019}]{Hennebelle2019}
{Hennebelle} P.,  {Inutsuka} S.-i.,  2019, \mn@doi [Frontiers in Astronomy and Space Sciences] {10.3389/fspas.2019.00005}, \href {https://ui.adsabs.harvard.edu/abs/2019FrASS...6....5H} {6, 5}

\bibitem[\protect\citeauthoryear{{Hernquist}}{{Hernquist}}{1990}]{Hernquist1990}
{Hernquist} L.,  1990, \mn@doi [\apj] {10.1086/168845}, \href {https://ui.adsabs.harvard.edu/abs/1990ApJ...356..359H} {356, 359}

\bibitem[\protect\citeauthoryear{{Heyer}, {Soler}  \& {Burkhart}}{{Heyer} et~al.}{2020}]{Heyer2020}
{Heyer} M.,  {Soler} J.~D.,   {Burkhart} B.,  2020, \mn@doi [\mnras] {10.1093/mnras/staa1760}, \href {https://ui.adsabs.harvard.edu/abs/2020MNRAS.496.4546H} {496, 4546}

\bibitem[\protect\citeauthoryear{{Hopkins}, {Quataert}  \& {Murray}}{{Hopkins} et~al.}{2012}]{Hopkins2012}
{Hopkins} P.~F.,  {Quataert} E.,   {Murray} N.,  2012, \mn@doi [\mnras] {10.1111/j.1365-2966.2012.20593.x}, \href {https://ui.adsabs.harvard.edu/abs/2012MNRAS.421.3522H} {421, 3522}

\bibitem[\protect\citeauthoryear{{Ib{\'a}{\~n}ez-Mej{\'\i}a}, {Mac Low}  \& {Klessen}}{{Ib{\'a}{\~n}ez-Mej{\'\i}a} et~al.}{2022}]{Ibanez2022}
{Ib{\'a}{\~n}ez-Mej{\'\i}a} J.~C.,  {Mac Low} M.-M.,   {Klessen} R.~S.,  2022, \mn@doi [\apj] {10.3847/1538-4357/ac3b58}, \href {https://ui.adsabs.harvard.edu/abs/2022ApJ...925..196I} {925, 196}

\bibitem[\protect\citeauthoryear{{Inoue} \& {Fukui}}{{Inoue} \& {Fukui}}{2013}]{Inoue2013}
{Inoue} T.,  {Fukui} Y.,  2013, \mn@doi [\apjl] {10.1088/2041-8205/774/2/L31}, \href {https://ui.adsabs.harvard.edu/abs/2013ApJ...774L..31I} {774, L31}

\bibitem[\protect\citeauthoryear{{Izquierdo} et~al.,}{{Izquierdo} et~al.}{2021}]{Izquierdo2021}
{Izquierdo} A.~F.,  et~al., 2021, \mn@doi [\mnras] {10.1093/mnras/staa3470}, \href {https://ui.adsabs.harvard.edu/abs/2021MNRAS.500.5268I} {500, 5268}

\bibitem[\protect\citeauthoryear{{Jeans}}{{Jeans}}{1902}]{Jeans1902}
{Jeans} J.~H.,  1902, \mn@doi [Philosophical Transactions of the Royal Society of London Series A] {10.1098/rsta.1902.0012}, \href {https://ui.adsabs.harvard.edu/abs/1902RSPTA.199....1J} {199, 1}

\bibitem[\protect\citeauthoryear{{Kim}, {Ostriker}  \& {Filippova}}{{Kim} et~al.}{2021}]{Kim2021}
{Kim} J.-G.,  {Ostriker} E.~C.,   {Filippova} N.,  2021, \mn@doi [\apj] {10.3847/1538-4357/abe934}, \href {https://ui.adsabs.harvard.edu/abs/2021ApJ...911..128K} {911, 128}

\bibitem[\protect\citeauthoryear{{Kim}, {Clark}, {Putman}  \& {Li}}{{Kim} et~al.}{2023}]{KimD2023}
{Kim} D.~A.,  {Clark} S.~E.,  {Putman} M.~E.,   {Li} L.,  2023, \mn@doi [\mnras] {10.1093/mnras/stad2792}, \href {https://ui.adsabs.harvard.edu/abs/2023MNRAS.526.4345K} {526, 4345}

\bibitem[\protect\citeauthoryear{{Klos}, {Bonnell}  \& {Smith}}{{Klos} et~al.}{2025}]{Klos2025}
{Klos} K.~S.,  {Bonnell} I.~A.,   {Smith} R.~J.,  2025, \mn@doi [\mnras] {10.1093/mnras/staf542}, \href {https://ui.adsabs.harvard.edu/abs/2025MNRAS.539.2307K} {539, 2307}

\bibitem[\protect\citeauthoryear{{Kong}, {Smith}, {Whitworth}  \& {Hamden}}{{Kong} et~al.}{2024}]{Kong2024}
{Kong} S.,  {Smith} R.~J.,  {Whitworth} D.,   {Hamden} E.~T.,  2024, \mn@doi [\apj] {10.3847/1538-4357/ad73a5}, \href {https://ui.adsabs.harvard.edu/abs/2024ApJ...975...97K} {975, 97}

\bibitem[\protect\citeauthoryear{{Korpi-Lagg}, {Mac Low}  \& {Gent}}{{Korpi-Lagg} et~al.}{2024}]{Korpi2024}
{Korpi-Lagg} M.~J.,  {Mac Low} M.-M.,   {Gent} F.~A.,  2024, \mn@doi [Living Reviews in Computational Astrophysics] {10.1007/s41115-024-00021-9}, \href {https://ui.adsabs.harvard.edu/abs/2024LRCA...10....3K} {10, 3}

\bibitem[\protect\citeauthoryear{{Lada} \& {Lada}}{{Lada} \& {Lada}}{2003}]{Lada2003}
{Lada} C.~J.,  {Lada} E.~A.,  2003, \araa, 41, 57

\bibitem[\protect\citeauthoryear{{Larson}}{{Larson}}{1981}]{Larson1981}
{Larson} R.~B.,  1981, \mn@doi [\mnras] {10.1093/mnras/194.4.809}, \href {https://ui.adsabs.harvard.edu/abs/1981MNRAS.194..809L} {194, 809}

\bibitem[\protect\citeauthoryear{{Li} \& {Zhao}}{{Li} \& {Zhao}}{2024}]{Li&Zhao2024}
{Li} G.-X.,  {Zhao} M.,  2024, \mn@doi [arXiv e-prints] {10.48550/arXiv.2409.02769}, \href {https://ui.adsabs.harvard.edu/abs/2024arXiv240902769L} {p. arXiv:2409.02769}

\bibitem[\protect\citeauthoryear{{Liu}, {Li}, {Staveley-Smith}, {Qian}, {Wong}  \& {Goldsmith}}{{Liu} et~al.}{2019}]{Liu2019}
{Liu} B.,  {Li} D.,  {Staveley-Smith} L.,  {Qian} L.,  {Wong} T.,   {Goldsmith} P.,  2019, \mn@doi [\apj] {10.3847/1538-4357/ab54cd}, \href {https://ui.adsabs.harvard.edu/abs/2019ApJ...887..242L} {887, 242}

\bibitem[\protect\citeauthoryear{{Mac Low} \& {Klessen}}{{Mac Low} \& {Klessen}}{2004}]{MacLow&Klessen2004}
{Mac Low} M.-M.,  {Klessen} R.~S.,  2004, \mn@doi [Reviews of Modern Physics] {10.1103/RevModPhys.76.125}, \href {https://ui.adsabs.harvard.edu/abs/2004RvMP...76..125M} {76, 125}

\bibitem[\protect\citeauthoryear{{Miyoshi} \& {Kusano}}{{Miyoshi} \& {Kusano}}{2005}]{Miyoshi2005}
{Miyoshi} T.,  {Kusano} K.,  2005, \mn@doi [Journal of Computational Physics] {10.1016/j.jcp.2005.02.017}, \href {https://ui.adsabs.harvard.edu/abs/2005JCoPh.208..315M} {208, 315}

\bibitem[\protect\citeauthoryear{{Myers} \& {Basu}}{{Myers} \& {Basu}}{2021}]{Myers2021}
{Myers} P.~C.,  {Basu} S.,  2021, \mn@doi [\apj] {10.3847/1538-4357/abf4c8}, \href {https://ui.adsabs.harvard.edu/abs/2021ApJ...917...35M} {917, 35}

\bibitem[\protect\citeauthoryear{{Ntormousi}, {Dawson}, {Hennebelle}  \& {Fierlinger}}{{Ntormousi} et~al.}{2017}]{Ntormousi2017}
{Ntormousi} E.,  {Dawson} J.~R.,  {Hennebelle} P.,   {Fierlinger} K.,  2017, \mn@doi [\aap] {10.1051/0004-6361/201629268}, \href {https://ui.adsabs.harvard.edu/abs/2017A&A...599A..94N} {599, A94}

\bibitem[\protect\citeauthoryear{{Padoan}, {Pan}, {Haugb{\o}lle}  \& {Nordlund}}{{Padoan} et~al.}{2016}]{Padoan2016}
{Padoan} P.,  {Pan} L.,  {Haugb{\o}lle} T.,   {Nordlund} {\r{A}}.,  2016, \apj, 822, 11

\bibitem[\protect\citeauthoryear{{Pakmor}, {Bauer}  \& {Springel}}{{Pakmor} et~al.}{2011}]{Pakmor2011}
{Pakmor} R.,  {Bauer} A.,   {Springel} V.,  2011, \mn@doi [\mnras] {10.1111/j.1365-2966.2011.19591.x}, \href {https://ui.adsabs.harvard.edu/abs/2011MNRAS.418.1392P} {418, 1392}

\bibitem[\protect\citeauthoryear{{Palau} et~al.,}{{Palau} et~al.}{2015}]{Palau2015}
{Palau} A.,  et~al., 2015, \mn@doi [\mnras] {10.1093/mnras/stv1834}, \href {https://ui.adsabs.harvard.edu/abs/2015MNRAS.453.3785P} {453, 3785}

\bibitem[\protect\citeauthoryear{{Palau} et~al.,}{{Palau} et~al.}{2021}]{Palau2021}
{Palau} A.,  et~al., 2021, \mn@doi [\apj] {10.3847/1538-4357/abee1e}, \href {https://ui.adsabs.harvard.edu/abs/2021ApJ...912..159P} {912, 159}

\bibitem[\protect\citeauthoryear{{Pattle}, {Fissel}, {Tahani}, {Liu}  \& {Ntormousi}}{{Pattle} et~al.}{2023}]{Pattle2023}
{Pattle} K.,  {Fissel} L.,  {Tahani} M.,  {Liu} T.,   {Ntormousi} E.,  2023, in {Inutsuka} S.,  {Aikawa} Y.,  {Muto} T.,  {Tomida} K.,   {Tamura} M.,  eds,  Astronomical Society of the Pacific Conference Series Vol. 534, Protostars and Planets VII. p.~193 (\mn@eprint {arXiv} {2203.11179}), \mn@doi{10.48550/arXiv.2203.11179}

\bibitem[\protect\citeauthoryear{{P{\'e}roux} \& {Howk}}{{P{\'e}roux} \& {Howk}}{2020}]{Peroux2020}
{P{\'e}roux} C.,  {Howk} J.~C.,  2020, \mn@doi [\araa] {10.1146/annurev-astro-021820-120014}, \href {https://ui.adsabs.harvard.edu/abs/2020ARA&A..58..363P} {58, 363}

\bibitem[\protect\citeauthoryear{{Planck Collaboration} et~al.,}{{Planck Collaboration} et~al.}{2016}]{PlanckCollab2016}
{Planck Collaboration} et~al., 2016, \mn@doi [\aap] {10.1051/0004-6361/201525896}, \href {https://ui.adsabs.harvard.edu/abs/2016A&A...586A.138P} {586, A138}

\bibitem[\protect\citeauthoryear{{Powell}, {Roe}, {Linde}, {Gombosi}  \& {De Zeeuw}}{{Powell} et~al.}{1999}]{Powell1999}
{Powell} K.~G.,  {Roe} P.~L.,  {Linde} T.~J.,  {Gombosi} T.~I.,   {De Zeeuw} D.~L.,  1999, \mn@doi [Journal of Computational Physics] {10.1006/jcph.1999.6299}, \href {https://ui.adsabs.harvard.edu/abs/1999JCoPh.154..284P} {154, 284}

\bibitem[\protect\citeauthoryear{{Ram{\'\i}rez-Galeano}, {Ballesteros-Paredes}, {Smith}, {Camacho}  \& {Zamora-Avil{\'e}s}}{{Ram{\'\i}rez-Galeano} et~al.}{2022}]{Ramirez2022}
{Ram{\'\i}rez-Galeano} L.,  {Ballesteros-Paredes} J.,  {Smith} R.~J.,  {Camacho} V.,   {Zamora-Avil{\'e}s} M.,  2022, \mnras, 515, 2822

\bibitem[\protect\citeauthoryear{{Salda{\~n}o}, {Rubio}, {Bolatto}, {Verdugo}, {Jameson}  \& {Leroy}}{{Salda{\~n}o} et~al.}{2023}]{Saldano2023}
{Salda{\~n}o} H.~P.,  {Rubio} M.,  {Bolatto} A.~D.,  {Verdugo} C.,  {Jameson} K.~E.,   {Leroy} A.~K.,  2023, \mn@doi [\aap] {10.1051/0004-6361/202142217}, \href {https://ui.adsabs.harvard.edu/abs/2023A&A...672A.153S} {672, A153}

\bibitem[\protect\citeauthoryear{{Schinnerer} \& {Leroy}}{{Schinnerer} \& {Leroy}}{2024}]{Schinnerer2024}
{Schinnerer} E.,  {Leroy} A.~K.,  2024, \mn@doi [\araa] {10.1146/annurev-astro-071221-052651}, \href {https://ui.adsabs.harvard.edu/abs/2024ARA&A..62..369S} {62, 369}

\bibitem[\protect\citeauthoryear{{Seifried}, {Walch}, {Weis}, {Reissl}, {Soler}, {Klessen}  \& {Joshi}}{{Seifried} et~al.}{2020}]{Seifried2020}
{Seifried} D.,  {Walch} S.,  {Weis} M.,  {Reissl} S.,  {Soler} J.~D.,  {Klessen} R.~S.,   {Joshi} P.~R.,  2020, \mn@doi [\mnras] {10.1093/mnras/staa2231}, \href {https://ui.adsabs.harvard.edu/abs/2020MNRAS.497.4196S} {497, 4196}

\bibitem[\protect\citeauthoryear{{Seta} \& {McClure-Griffiths}}{{Seta} \& {McClure-Griffiths}}{2025}]{Seta2025}
{Seta} A.,  {McClure-Griffiths} N.~M.,  2025, \mn@doi [\mnras] {10.1093/mnras/staf520}, \href {https://ui.adsabs.harvard.edu/abs/2025MNRAS.539.1024S} {539, 1024}

\bibitem[\protect\citeauthoryear{{Smith}, {Longmore}  \& {Bonnell}}{{Smith} et~al.}{2009}]{Smith2009b}
{Smith} R.~J.,  {Longmore} S.,   {Bonnell} I.,  2009, \mnras, 400, 1775

\bibitem[\protect\citeauthoryear{{Smith}, {Glover}, {Clark}, {Klessen}  \& {Springel}}{{Smith} et~al.}{2014}]{Smith2014}
{Smith} R.~J.,  {Glover} S. C.~O.,  {Clark} P.~C.,  {Klessen} R.~S.,   {Springel} V.,  2014, \mn@doi [\mnras] {10.1093/mnras/stu616}, \href {https://ui.adsabs.harvard.edu/abs/2014MNRAS.441.1628S} {441, 1628}

\bibitem[\protect\citeauthoryear{{Smith} et~al.,}{{Smith} et~al.}{2020}]{Smith2020}
{Smith} R.~J.,  et~al., 2020, \mnras, 492, 1594

\bibitem[\protect\citeauthoryear{{Smith} et~al.,}{{Smith} et~al.}{2023}]{Smith2023}
{Smith} R.~J.,  et~al., 2023, \mn@doi [\mnras] {10.1093/mnras/stad1537}, \href {https://ui.adsabs.harvard.edu/abs/2023MNRAS.524..873S} {524, 873}

\bibitem[\protect\citeauthoryear{{Soler}}{{Soler}}{2019}]{Soler2019}
{Soler} J.~D.,  2019, \mn@doi [\aap] {10.1051/0004-6361/201935779}, \href {https://ui.adsabs.harvard.edu/abs/2019A&A...629A..96S} {629, A96}

\bibitem[\protect\citeauthoryear{{Soler} \& {Hennebelle}}{{Soler} \& {Hennebelle}}{2017}]{Soler2017}
{Soler} J.~D.,  {Hennebelle} P.,  2017, \mn@doi [\aap] {10.1051/0004-6361/201731049}, \href {https://ui.adsabs.harvard.edu/abs/2017A&A...607A...2S} {607, A2}

\bibitem[\protect\citeauthoryear{{Soler}, {Hennebelle}, {Martin}, {Miville-Desch{\^e}nes}, {Netterfield}  \& {Fissel}}{{Soler} et~al.}{2013}]{Soler2013}
{Soler} J.~D.,  {Hennebelle} P.,  {Martin} P.~G.,  {Miville-Desch{\^e}nes} M.~A.,  {Netterfield} C.~B.,   {Fissel} L.~M.,  2013, \mn@doi [\apj] {10.1088/0004-637X/774/2/128}, \href {https://ui.adsabs.harvard.edu/abs/2013ApJ...774..128S} {774, 128}

\bibitem[\protect\citeauthoryear{{Sormani}, {Tre{\ss}}, {Klessen}  \& {Glover}}{{Sormani} et~al.}{2017}]{Sormani2017}
{Sormani} M.~C.,  {Tre{\ss}} R.~G.,  {Klessen} R.~S.,   {Glover} S. C.~O.,  2017, \mn@doi [\mnras] {10.1093/mnras/stw3205}, \href {https://ui.adsabs.harvard.edu/abs/2017MNRAS.466..407S} {466, 407}

\bibitem[\protect\citeauthoryear{{Springel}}{{Springel}}{2010}]{Springel2010}
{Springel} V.,  2010, \mn@doi [\mnras] {10.1111/j.1365-2966.2009.15715.x}, \href {https://ui.adsabs.harvard.edu/abs/2010MNRAS.401..791S} {401, 791}

\bibitem[\protect\citeauthoryear{{Syed} et~al.,}{{Syed} et~al.}{2020}]{Syed2020}
{Syed} J.,  et~al., 2020, \mn@doi [\aap] {10.1051/0004-6361/202038449}, \href {https://ui.adsabs.harvard.edu/abs/2020A&A...642A..68S} {642, A68}

\bibitem[\protect\citeauthoryear{{Tress}, {Smith}, {Sormani}, {Glover}, {Klessen}, {Mac Low}  \& {Clark}}{{Tress} et~al.}{2020}]{Tress2020}
{Tress} R.~G.,  {Smith} R.~J.,  {Sormani} M.~C.,  {Glover} S. C.~O.,  {Klessen} R.~S.,  {Mac Low} M.-M.,   {Clark} P.~C.,  2020, \mn@doi [\mnras] {10.1093/mnras/stz3600}, \href {https://ui.adsabs.harvard.edu/abs/2020MNRAS.492.2973T} {492, 2973}

\bibitem[\protect\citeauthoryear{{Truelove}, {Klein}, {McKee}, {Holliman}, {Howell}  \& {Greenough}}{{Truelove} et~al.}{1997}]{Truelove1997}
{Truelove} J.~K.,  {Klein} R.~I.,  {McKee} C.~F.,  {Holliman} John~H. I.,  {Howell} L.~H.,   {Greenough} J.~A.,  1997, \mn@doi [\apjl] {10.1086/310975}, \href {https://ui.adsabs.harvard.edu/abs/1997ApJ...489L.179T} {489, L179}

\bibitem[\protect\citeauthoryear{{V{\'a}zquez-Semadeni}, {Palau}, {Ballesteros-Paredes}, {G{\'o}mez}  \& {Zamora-Avil{\'e}s}}{{V{\'a}zquez-Semadeni} et~al.}{2019}]{Vazquez2019}
{V{\'a}zquez-Semadeni} E.,  {Palau} A.,  {Ballesteros-Paredes} J.,  {G{\'o}mez} G.~C.,   {Zamora-Avil{\'e}s} M.,  2019, \mnras, 490, 3061

\bibitem[\protect\citeauthoryear{{Weis}, {Walch}, {Seifried}  \& {Ganguly}}{{Weis} et~al.}{2024}]{Weis2024}
{Weis} M.,  {Walch} S.,  {Seifried} D.,   {Ganguly} S.,  2024, \mn@doi [\mnras] {10.1093/mnras/stae1518}, \href {https://ui.adsabs.harvard.edu/abs/2024MNRAS.532.1262W} {532, 1262}

\bibitem[\protect\citeauthoryear{Whitworth, Smith, Klessen, Mac Low, Glover, Tress, Pakmor  \& Soler}{Whitworth et~al.}{2023}]{Whitworth2023}
Whitworth D.~J.,  Smith R.~J.,  Klessen R.~S.,  Mac Low M.-M.,  Glover S. C.~O.,  Tress R.,  Pakmor R.,   Soler J.~D.,  2023, \mn@doi [Monthly Notices of the Royal Astronomical Society] {10.1093/mnras/stad105}, 520, 89

\bibitem[\protect\citeauthoryear{{Whitworth} et~al.,}{{Whitworth} et~al.}{2025a}]{Whitworth2025a}
{Whitworth} D.~J.,  et~al., 2025a, \mn@doi [\mnras] {10.1093/mnras/stae2759}, \href {https://ui.adsabs.harvard.edu/abs/2025MNRAS.536.2936W} {536, 2936}

\bibitem[\protect\citeauthoryear{{Whitworth} et~al.,}{{Whitworth} et~al.}{2025b}]{Whitworth2025b}
{Whitworth} D.~J.,  et~al., 2025b, \mn@doi [\mnras] {10.1093/mnras/staf901}, \href {https://ui.adsabs.harvard.edu/abs/2025MNRAS.540.2762W} {540, 2762}

\bibitem[\protect\citeauthoryear{{Wolfire}, {Hollenbach}, {McKee}, {Tielens}  \& {Bakes}}{{Wolfire} et~al.}{1995}]{Wolfire1995}
{Wolfire} M.~G.,  {Hollenbach} D.,  {McKee} C.~F.,  {Tielens} A.~G.~G.~M.,   {Bakes} E.~L.~O.,  1995, \mn@doi [\apj] {10.1086/175510}, \href {https://ui.adsabs.harvard.edu/abs/1995ApJ...443..152W} {443, 152}

\bibitem[\protect\citeauthoryear{{Wolfire}, {McKee}, {Hollenbach}  \& {Tielens}}{{Wolfire} et~al.}{2003}]{Wolfire2003}
{Wolfire} M.~G.,  {McKee} C.~F.,  {Hollenbach} D.,   {Tielens} A.~G.~G.~M.,  2003, \mn@doi [\apj] {10.1086/368016}, \href {https://ui.adsabs.harvard.edu/abs/2003ApJ...587..278W} {587, 278}

\bibitem[\protect\citeauthoryear{{Xu}, {Ji}  \& {Lazarian}}{{Xu} et~al.}{2019}]{Xu2019}
{Xu} S.,  {Ji} S.,   {Lazarian} A.,  2019, \mn@doi [\apj] {10.3847/1538-4357/ab21be}, \href {https://ui.adsabs.harvard.edu/abs/2019ApJ...878..157X} {878, 157}

\makeatother
\end{thebibliography}




\appendix

\section{Density cut for Nearest Neighbour search}
\label{append:nn_cut}

To accurately compute a velocity dispersion from the nearest neighbour search, only cells with a similar density (and thus similar state) need to be used. This is because the velocity of the gas will depend on its state. A nearest neighbour search by itself can only find all cells within a given radius of a target cell, and so we must also include a density cut. Such a cut would be of the form: n$_{\rm cell}$ > 1/$f$ $\times$ n$_{\rm target}$ and n$_{\rm cell}$ < $f$ $\times$ n$_{\rm target}$. Where n$_{\rm target}$ is the number density of the target cell, n$_{\rm cell}$ is the  number density of surrounding cells and f is the density cut factor to determine. To find the most suitable $f$, the nearest neighbour search and velocity dispersion calculation (as described in Section~\ref{vel_dis}) was performed on the low Z model for values of f ranging from two to twenty. Since doing so results in millions of values, the mean of the velocity dispersions was used for analysis. Additionally, this was performed on the low Z model since it had the fewest cells and thus ran the fastest. Due to the statistical similarities between this model and the solar model the same density cut value can be used. The result is summarised in Figure~\ref{fig:density_cut}.

\begin{figure}
    \centering
    \includegraphics[width=\columnwidth]{Figures/lowZ_plots/nn_density_cut.png}
    \caption{The effect on the mean nearest neighbour velocity dispersion from the choice of density cut factor, $f$.}
    \label{fig:density_cut}
\end{figure}
 
The two most important features are at either end. On the left, we see that choosing a factor of two results in a much lower average velocity dispersion compared to other cuts, suggesting this condition is too strict. On the other end, the plot flattens out, beyond a factor of eleven. This suggests most cells are being included with these cuts and likely overestimating the velocity dispersion. Thus, the appropriate cut to make would be between a factor of five to ten. Whilst the choice seems somewhat arbitrary, the difference in results between choosing five or ten is minimal. When performing the energy balance methods to obtain how much mass is dominated by turbulence, a density cut of ten is less than one percentage point greater than choosing a factor of five. 

Due to this minimal difference, we choose the middle of this range, f = 8.

\bsp	
\label{lastpage}
\end{document}